\documentclass[superscriptaddress,aps,pra,twocolumn]{revtex4}
\usepackage{bm}
\usepackage{ulem}
\usepackage{epsfig}
\usepackage{graphicx}
\usepackage{amssymb,amsmath,amsbsy,amsgen,amsfonts}    
\usepackage{dcolumn}
\usepackage{amsthm}
\usepackage{mathrsfs}
\usepackage{latexsym}
\usepackage{array}
\usepackage{color}
\usepackage{amstext}
\allowdisplaybreaks[1]
\usepackage{txfonts}
\usepackage{pifont}
\usepackage{braket}
\usepackage{epstopdf} 

\newcommand{\im}{{\rm i}}

\newcommand{\ketbra}[2]{|#1\rangle \langle#2|}

\newcommand{\be}{\begin{equation}}
\newcommand{\ee}{\end{equation}}
\newcommand{\ba}{\begin{array}}
\newcommand{\ea}{\end{array}}
\newcommand{\bqa}{\begin{eqnarray}}
\newcommand{\eqa}{\end{eqnarray}}

\newcommand{\aalpha}{{\hat{a}}_{\alpha}}
\newcommand{\aalphad}{{\hat{a}^\dagger}_{\alpha}}
\newcommand{\axi}{{\hat{a}}_{\xi}}
\newcommand{\axid}{\hat{a}_\xi^\dagger}
\newcommand{\at}[1][]{\hat{a}(t_{#1})}
\newcommand{\atd}[1][]{\hat{a}^\dagger(t_{#1})}

\newcommand{\coh}{|\{\alpha\}\rangle}

\newcommand{\packet}{|\{\alpha\},1_\xi\rangle}

\DeclareSymbolFont{symbols}{OMS}{cmsy}{m}{n}

\begin{document}

\title{Photon-added coherent states using the continuous-mode formalism}

\author{J. T. Francis}
\affiliation{Laser Research Institute, Department of Physics, Stellenbosch University, Private Bag X1, Matieland 7602, South Africa}
\author{M. S. Tame}
\affiliation{Laser Research Institute, Department of Physics, Stellenbosch University, Private Bag X1, Matieland 7602, South Africa}

\date{\today}

\begin{abstract}
The addition of a photon into the same mode as a coherent state produces a nonclassical state that has interesting features, including quadrature squeezing and a sub-Poissonian photon-number distribution. The squeezed nature of photon-added coherent (PAC) states potentially offers an advantage in quantum sensing applications. Previous theoretical works have employed a single-mode treatment of PAC states. Here, we use a continuous-mode approach that allows us to model PAC state pulses. We study the properties of a single-photon and coherent state wavepacket superimposed with variable temporal and spectral overlap. We show that, even without perfect overlap, the state exhibits a sub-Poissonian number distribution, second-order quantum correlations and quadrature squeezing for a weak coherent state. We also include propagation loss in waveguides and study how the fidelity and other properties of PAC state pulses are affected. 
\end{abstract}

\maketitle

\section{Introduction}

Nonclassical states are important resources for quantum information processing and probing fundamental properties of quantum mechanics~\cite{Dodonov02,Braunstein05}. A wide range of states have been studied in the literature, most notably in optical-based systems, including Fock states~\cite{Waks06}, displaced Fock states~\cite{Satya85,Ziesel13} and different types of squeezed states~\cite{Lvovsky15}. In particular, photon-added coherent (PAC) states~\cite{Agarwal91,Agarwal92}, where a photon is added to the same mode as a coherent state, have received much attention~\cite{Sivakumar99,Sivakumar00}. They have applications in quantum sensing~\cite{Braun14,Gard16,Schnabel17} and helping to develop security protocols in quantum key distribution~\cite{Loepp06,Assche06,Barnett06,Barnett18}. Furthermore, the process of adding (and subtracting) a photon in an optical field has interesting physical consequences~\cite{Kim08}, enabling quantum state engineering~\cite{Fiurasek09} and the probing of quantum features, such as bosonic commutation relations~\cite{Parigi07} and quantum thermodynamics~\cite{Vidrighin16}. In recent years, theoretical studies have investigated the generation of PAC states in cavity and ion-trap systems~\cite{RamosPrieto14}, and their creation using photon-subtracted states~\cite{Mojaveri14}, amplification methods~\cite{Shringapure19} and nonlinear optics~\cite{Kalamidas08,Li07}. Studies have also investigated their entanglement properties~\cite{Dominguez16,Ren19}, robustness to noise and dissipation~\cite{Dominguez16,Hu09}, in addition to their statistics~\cite{Barnett18}, practical characterisation~\cite{Filippov13} and generalisation to more complex structured states~\cite{HongChun10,Sivakumar13}. On the experimental side, studies have investigated the generation of PAC states by parametric down-conversion~\cite{Zavatta04}, as well as the characterisation of properties, such as photon statistics and the Wigner function~\cite{Zavatta05}, and degree of non-Gaussianity~\cite{Barbieri10}. 

Although considerable progress has been made in the development of PAC states, an important issue is that the theoretical works carried out so far use a single-mode description, while in experiments pulsed light is used, which naturally requires a continuous (temporal) mode description~\cite{Blow90,Loudon00}. The results observed in pulsed experiments are roughly in line with the single-mode theory, however, the impact of temporal and spectral wavepacket imperfections on the properties of PAC states cannot be predicted from a single-mode picture. For example, it is not possible to predict how a mismatch in the pulse duration of the single photon and coherent state wavepackets, when added together, affects the sub-Poissonian behaviour and quadrature squeezing. This may have an adverse effect on the performance of the generated PAC states in quantum sensing and other applications.

In this work, we use a continuous-mode formalism to show how the properties of PAC states are affected by timing and bandwidth imperfections, as well as loss from propagation in waveguides. We study the photon-number distribution, second-order correlations, quadrature squeezing and fidelity of pulsed PAC states. We find that PAC states are reasonably robust to temporal and spectral mismatch, as well as propagation loss. The results of the work may help in the further development of experimental schemes for PAC state generation and their use in quantum information applications.

In Section II we introduce the model for the work and some preliminary details, including some mathematical relations that will be used throughout the study. In Section III we investigate the photon statistics of continuous-mode PAC states, including photon-number distribution and second-order correlations. In Section IV we study quadratures and squeezing, and in Section V we derive an expression for the fidelity. Finally, in Section VI we summarize our findings.

\begin{figure*}[t]
\centering
\includegraphics[width=17.6cm]{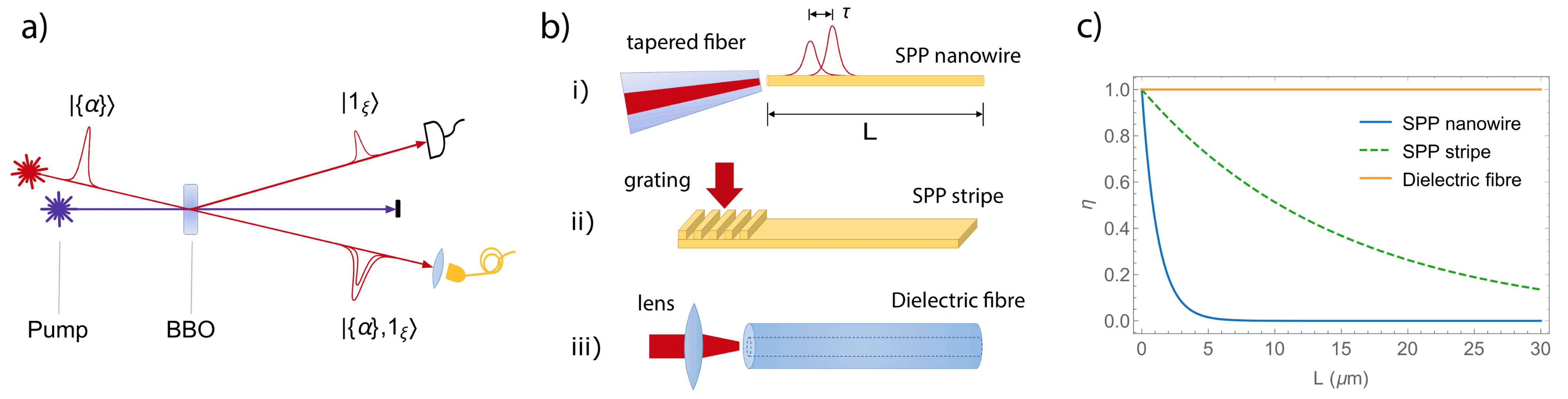}
\caption{Generation and propagation of continuous-mode photon-added coherent (PAC) state pulses. (a) A scheme for generating a PAC state using an optical parametric amplifier with a Beta Barium Borate (BBO) crystal. A coherent state is sent into the signal mode and a down-converted photon is emitted into the same spatial mode via stimulated emission. A second photon is emitted into the idler mode and used to herald the successful photon addition. Ideally, the stimulated photon is in phase with the coherent state and has the same temporal shape. However, temporal properties of the pump pulse may affect the emission time, duration and shape of the added photon wavepacket. (b) Propagation of the continuous-mode PAC state pulses in three example waveguides: (i) surface plasmon polariton (SPP) nanowire, (ii) SPP stripe and (iii) dielectric fibre. In all waveguides a perfect coupling from free space is assumed. Imperfect coupling can be included in an overall loss factor~\cite{Tame08}. (c) Loss incurred, $\eta$, for a given propagation distance $L$ in the waveguides considered in (b). Here, the loss factor $\eta=e^{-k_\im L}$, with $k_\im$ set as $1/1.2$, $1/15$ and $1/10^{10}$ as examples for light at optical wavelengths in the respective waveguides.}
\label{fig1} 
\end{figure*}

\section{Model and Preliminaries}
\subsection{Photon-added coherent state}
A salient feature of the continuous mode formalism is the use of frequency dependent photon annihilation and creation operators $\hat{a}(\omega)$ and  $\hat{a}^\dagger(\omega)$, respectively. The Fourier transforms of these operators give the instantaneous operators
\begin{equation}
\hat{a}(t) = \frac{1}{\sqrt{2\pi}} \int d\omega \hat{a}(\omega)\mathrm{e}^{-\im \omega t}
\end{equation}
and its Hermitian conjugate $\hat{a}^\dagger(t)$. These operators obey the commutation relation
\begin{equation}
[\hat{a}(t),\hat{a}^\dagger(t')] = \delta(t' - t).
\label{eqn:atcomm}
\end{equation}
Using $\hat{a}^\dagger(t)$ a photon wavepacket creation operator may be defined as 
\begin{equation}
\axid = \int dt \xi(t) \atd,
\label{eqn:axid}
\end{equation}
where $\xi(t)$ is the wavepacket amplitude.

A continuous-mode PAC state in a single spatial mode is constructed by the action of the photon wavepacket creation operator, $\axid$, with pulse profile $\xi(t + \tau)$, on a spatial mode containing a pulsed coherent state, $\coh$, with pulse profile $\alpha(t)$~\cite{Blow90,Loudon00}. The parameter $\tau$ enables the single photon and coherent state pulses to be offset in time. The resulting state,
	\begin{equation}
		\packet = |N|^{1/2}~\axid~\coh,
		\label{eqn:pacs}
	\end{equation}
is renormalised by the factor $|N|^{1/2}$. The normalisation constant is obtained from the relation $\braket{1_{\xi},\{\alpha\}|\{\alpha\},1_\xi}=1$, by substituting in Eq.~\eqref{eqn:pacs} and writing the operators in normal-order using the commutation relation $[\axi,\axid] = 1$. The following equation is also used,
	\begin{equation}
		\axi\coh  = \sigma(\tau)\coh,
		\label{eqn:axieig}
	\end{equation}
which is obtained from the relations $\axi = \int dt \xi^*(t+\tau) \at$ and $\at \ket{\{\alpha\}} = \alpha(t)\ket{\{\alpha\}}$ (see Appendix \ref{appendix: CMF}). The quantity $\sigma(\tau)$, given by
\begin{equation}
	\sigma(\tau) = \int dt~\xi^*(t + \tau)\alpha(t),
	\label{eqn:sigma}
\end{equation} 
is the cross-correlation of the single-photon and coherent state pulse profiles, and serves as a measure of the overlap between the pulses. The above steps lead to the normalization
	\begin{equation}
	|N(\tau)| = \Big(1 + |\sigma(\tau)|^2\Big)^{-1}.
	\end{equation}
In the limit of perfect overlap, {\it i.e.} $\tau=0$ and $\xi(t)=\alpha(t)/\sqrt{n_{\alpha}}$, where $n_{\alpha}$ is the mean photon number of the coherent state, we have $|\sigma(\tau)|^2=n_{\alpha}$ and a normalization $N=(1+n_{\alpha})^{-1}$, which recovers the well-known single-mode result~\cite{Agarwal91}. 

To the best of our knowledge, this continuous-mode version of a PAC state has not been considered before. While similar to the single-mode case on which it is based, its construction is not obvious as it involves an overlap between a single photon and a coherent state wavepacket. This naturally leads to the possibility of temporal and bandwidth mismatch in the state, which may occur depending on the physical scenario.
	
As an example, the continuous-mode PAC state introduced above can be produced using the process shown in Fig.~\ref{fig1}~(a), where a coherent state is fed into the signal spatial mode of an optical parametric amplifier (OPA) and a single down-converted photon is emitted into the same mode via stimulated emission, with another photon emitted into the idler mode and used to herald the successful photon addition~\cite{Zavatta04,Zavatta05}. Under ideal conditions the stimulated photon in the signal mode produced by the ensemble of atoms in the OPA will be in phase with the coherent state and have the same temporal shape~\cite{Vahala93}. However, the arrival time, duration and shape of the pump pulse entering the OPA (BBO crystal) will affect the emission time, duration and shape of the added photon wavepacket. This may cause a mismatch in the overlap between it and the coherent state. For instance, the pump pulse could have a time duration narrower than the coherent state, and potentially even an offset in its arrival time at the crystal. 

Another example is in the stimulated emission from a single excited atom by a coherent state~\cite{Sivakumar99}, under the assumption of a weak intensity, well-defined spatial mode and long enough pulse profile~\cite{Vahala93,Fischer18}. Due to the finite lifetime of the atom's excited state, a pump pulse must be used to place the atom into this state close in time to the arrival of the coherent state wavepacket. If the timing of the pump is not exact, it will affect the time at which the atom can produce a stimulated photon in the same mode as the coherent state. For instance, the atom might only be put into the excited state by the pump after part of the coherent state pulse has already passed, causing a mismatch in the overlap between it and the coherent state.

Other scenarios for generating PAC states can be considered with different timing and bandwidth imperfections~\cite{Sivakumar00}. For instance, when adding a spontaneously emitted photon in phase with a coherent state~\cite{Resch02,Rarity05}. To keep the model general we consider the case where the coherent state and the single-photon wavepackets have arbitrary time profiles with time durations and pulse centre times varied as independent parameters. As mentioned earlier, the single photon pulse in our model has a complex wavepacket amplitude, or profile, $\xi(t+\tau)$. This profile has a bandwidth $\Omega_1$, phase $\theta_1(t + \tau) = \omega_0(t + \tau)$ and pulse centre shifted in time by $\tau$ with respect to that of the coherent state. The coherent state, with profile $\alpha(t)$, has a mean photon number of $n_\alpha$, a bandwidth of $\Omega$, and a phase $\theta(t) = \omega_0 t$. The peak of the coherent state pulse passes the coordinate origin $z=0$ of the single spatial mode that it occupies at time $t_0=0$. For simplicity, we limit our study to pulses of the same central frequency $\omega_0$ and assume that the coherent state frequency can be tuned to closely match that of the single photon~\cite{Zavatta04}. All the results shown are for the case of Gaussian pulses, however the theory is applicable to more general pulse profiles.

In the above model, we have described PAC states using the continuous-mode formalism of Refs.~\cite{Blow90,Loudon00}, as it is most appropriate for narrowband wavepackets. Brief descriptions of number state and coherent state pulses are provided in Appendix~\ref{appendix: CMF} for the interested reader, with minor details required for some of the derivations given in the remainder of the work. 

\subsection{Propagation}

Finally, we describe the model used for the propagation of continuous-mode PAC states in various types of waveguides, such as those shown in Fig.~\ref{fig1}~(b). As example waveguides, we consider a surface plasmon polariton (SPP) nanowire~\cite{Akimov07}, SPP stripe~\cite{DiMartino12} and dielectric fibre~\cite{OBrien05}, all with varying amounts of loss incurred for a given propagation distance, as described by the function $\eta$ shown in Fig.~\ref{fig1}~(c).

An input state propagating in a waveguide with a complex dispersion relation $k(\omega) = k_r(\omega) + \im k_i(\omega)$ will undergo a phase-shift and attenuation due to the real and imaginary parts of $k(\omega)$, respectively. The travelling-wave attenuation model of Ref.~\cite{Jeffers93} enables a description of such propagation using an effective beamsplitter system. The input-output relations are
	\begin{equation}
		\hat{a}_{L}(t) = \eta^{\frac{1}{2}}(L)~\at[r] + \im(1-\eta(L))^{\frac{1}{2}}~\hat{v}(t)
		\label{eqn:aL}
	\end{equation}
and	
	\begin{equation}
	\hat{v}_{L}(t) = \eta^{\frac{1}{2}}(L)~\hat{v}(t) + \im(1-\eta(L))^{\frac{1}{2}}~\hat{a}(t_r).
	\label{eqn:vL}
	\end{equation}
The operators $\hat{a}(t)$ and $\hat{v}(t)$ represent the initial single guided and environment (bath) modes respectively, while $\hat{a}_L(t)$ and $\hat{v}_L(t)$ are correspondingly the modes after the state propagates a distance $L$. A retarded time, $t_r$, takes into account the time for a state to propagate the distance $L$ and the loss factor, $\eta(L)$, represents the amount of loss experienced during the propagation. In the next sections, loss is introduced into observables by substituting $\hat{a}(t) \rightarrow \hat{a}_L(t)$, $\hat{v}(t) \rightarrow \hat{v}_L(t)$, and similary for their Hermitian conjugates, with the environment taken to be in the vacuum state initially and traced out finally. 

The loss factor, $\eta(L) = \mathrm{e}^{2\im k_r(\omega)L}\mathrm{e}^{-k_i(\omega)L}$, is approximately 
	\begin{equation}
	\eta(L) \simeq \mathrm{e}^{2\im [k_r(\omega_0)L - \omega_0 L/v_g(\omega_0)]}~\mathrm{e}^{-k_i(\omega_0)L}
	\label{eqn:etaL}
	\end{equation}
for a narrowband input state. The amplitude and phase of $\eta(L)$ are identified as $|\eta(L)| = \exp(-k_i(\omega_0)L)$ and $\varphi_\eta = 2[k_r(\omega_0)L - \omega_0 L/v_g(\omega_0)]$, respectively. The retarded time is
	\begin{equation}
		t_r = t - \frac{L}{v_g(\omega_0)},
		\label{eqn:tr}
	\end{equation}
where $v_g(\omega_0) = d\omega/dk|_{\omega_0}$ is the wavepacket group velocity. 

\section{Photon Statistics} 
We start our investigation of continuous-mode PAC states by studying their photon statistics. In particular, the photon number distribution and second-order correlation function are considered, and we study the impact of temporal mismatch and loss on these statistical properties. Both are experimentally relevant for the characterization of PAC states~\cite{Zavatta04,Zavatta05,Barbieri10}.

\subsection{Photon number distribution}
The first property we investigate is the photon number distribution $P_n$, its mean $\braket{n}$, and variance $(\Delta n)^2$. We begin by deriving the expression for $P_n$ and then describe its modification when propagation loss is included. These expressions are then evaluated for PAC states with arbitrary temporal profiles for the combined single-photon and coherent state.

Due to the composite nature (ambiguity in the pulse profile) of PAC states, we first calculate the probability density $P_n\big(\{t_i\}_{i=1}^{n}\big)$ by projecting out the transient number state defined as
\begin{equation}
\ket{\{1_{t_i}\}_{i=1}^{n}} = \frac{1}{\sqrt{\cal N}} \prod_{i=1}^{n} \atd[i]\ket{0},
\label{eqn:nt}	
\end{equation}
(see Appendix~\ref{appendix: CMF} for further details) giving,
	\begin{equation}
	P_n\big(\{t_i\}_{i=1}^{n}\big) = |\braket{\{1_{t_i}\}_{i=1}^{n}|\psi}|^2.
	\label{eqn:Pdensity}
	\end{equation}
The projection selects out the probability amplitude of the $n$-photon eigenstate in the state $\ket{\psi}$, which may be in a superposition of different photon numbers. This approach allows us to avoid specifying the profile of the state $\ket{\psi}$. Integrating the probability density then gives the probability distribution,
	\begin{equation}
		P_n = \int dt_1\cdots dt_n~P_n\big(\{t_i\}_{i=1}^{n}\big).
		\label{eqn:Pdist}
	\end{equation}
Eq.~\eqref{eqn:Pdist} is valid for photon numbers $n \geq 1$, as Eq.~\eqref{eqn:Pdensity} yields the  probability $P_0$ (and not the density) when the vacuum state $\ket{0}$ is projected out. For consistency, we restrict Eq.~\eqref{eqn:Pdensity} to $n\geq1$, and treat the vacuum state separately using
	\begin{equation}
		P_0 = 1 - \sum_{n=1}^{\infty} P_n.
		\label{eqn:P0}
	\end{equation}

Introducing loss during propagation requires that we project out the guided mode state $\ket{\{1_{t_i}\}_{i=1}^n}_g$, as well as trace out the environment states $\ket{\{1_{t_i}\}_{i=n+1}^{m}}_e \forall~m\geq n$. As an example, we consider a guided mode state having $m \geq n$ photons present, $m-n$ of which can be lost to the environment. In the Heisenberg picture the projectors evolve, while the state remains the same. We denote the initial state as $\rho=\ketbra{\Psi}{\Psi}$, where $\ket{\Psi}=\ket{\psi}_g\ket{0}_e$, and the projectors as $\hat{P}_{g,L}=\ketbra{\{ 1_{t_i}\}_{i=1}^{n}}{\{ 1_{t_i}\}_{i=1}^{n}}$ and $\hat{P}_{e,L}=\ketbra{\{ 1_{t_i}\}_{i=n+1}^{m}}{\{ 1_{t_i}\}_{i=n+1}^{m}}$, which are constructed using the definition of the transient number state in Appendix~\ref{appendix: CMF} and the replacement $\hat{a}(t)\to\hat{a}_L(t)$ and $\hat{v}(t)\to\hat{v}_L(t)$ to account for the evolution (propagation). The joint probability density of having $n$ photons in the guided mode and $m-n$ in the environment after a propagation distance $L$ is given by
\be
P_{n,m-n}\big(\{ t_i\}_{i=1}^{m},L\big)={\rm Tr}\big[ \hat{P}_{g,L} (\{ t_i\}_{i=1}^{n}) \otimes \hat{P}_{e,L} (\{ t_i\}_{i=n+1}^{m})\rho \big].
\ee
We then have the expression
\bqa
&&	P_{n,m-n}\big(\{t_i\}_{i=1}^m,L\big)  =  \frac{1}{n!(m-n)!} \times \nonumber \\
&& \qquad \qquad \qquad \big|\bra{0}_e\bra{0}_g\prod_{j=n+1}^{m}\hat{v}_L(t_j)\prod_{i=1}^{n}\hat{a}_L(t_i)\ket{\psi}_g\ket{0}_e\big|^2, \nonumber
\label{eqn:PnmtL}
\eqa
in which the factorials arise from the normalisation of the transient number states (see Appendix~\ref{appendix: CMF}). Next, the expressions for $\hat{a}_L(t)$ and $\hat{v}_L(t)$ from Eqs. (\ref{eqn:aL}) and (\ref{eqn:vL}) are used, taking into account that the environment is initially in a vacuum so that the expectation value of all terms with $\hat{v}$ vanishes. The photon number probability density is then
	\begin{equation}
		P_{n,m-n}\big(\{t_i\}_{i=1}^m,L\big) = {|\eta(L)|}^n{|1-\eta(L)|}^{m-n}~\binom{m}{n}~	P_m\big(\{t_{r_i}\}_{i=1}^m\big).
		\label{eqn:PnmdensityL}
	\end{equation}
Integrating with respect to all times gives the probability distribution
 \begin{equation}
 	P_{n,m-n}\big(L\big) = {|\eta(L)|}^n{|1-\eta(L)|}^{m-n}~\binom{m}{n}~P_m.
 	\label{eqn:PnmdistL}
 \end{equation}
In general the initial guided state may be a superposition of number states, in which case the environment will evolve into a superposition state. This necessitates tracing out all the environment states. Thus, the probability of exactly $n$ photons remaining in the guided mode after propagating a distance $L$ is then simply Eq.~\eqref{eqn:PnmdistL} summed for all $m \geq n$; 
\begin{equation}
	P_n(L) = 
		{|\eta(L)|}^n~\sum_{m=n}^{\infty}{|1-\eta(L)|}^{m-n}~\binom{m}{n}~P_m.
	\label{eqn:PnL}
\end{equation}
Here, $P_m\big(\{t_{r_i}\}_{i=1}^m\big)$ and $P_m$ are the lossless probability density (retarded) and probability distribution defined in Eqs.~\eqref{eqn:Pdensity} and \eqref{eqn:Pdist}, respectively. The above formula in Eq.~(\ref{eqn:PnL}) is a continuous-mode generalisation of the single-mode result that applies a Bernoulli transformation to a state's probability distribution in order to account for loss~\cite{Gardiner00}. 

We are now in a position to apply the continuous-mode probability density and its associated probability distribution to the PAC states we consider. We allow $n=0,1,2,\cdots$ in Eq.~\eqref{eqn:PnL}, but make the restriction that $0<|\eta(L)|<1$. The boundaries of $|\eta(L)|$ are excluded due the occurrence of $0^0$ when $n = 0$ and $n = m$. When $|\eta(L)|=0$ we have $P_n(L) = \delta_{0,n}$ for any initial state, since all photons will have been lost. When $|\eta(L)| = 1$, the lossless distribution derived for a specific initial state from Eq.~\eqref{eqn:Pdensity} is used.
\begin{figure*}[t]
\centering
\includegraphics[width=17.8cm]{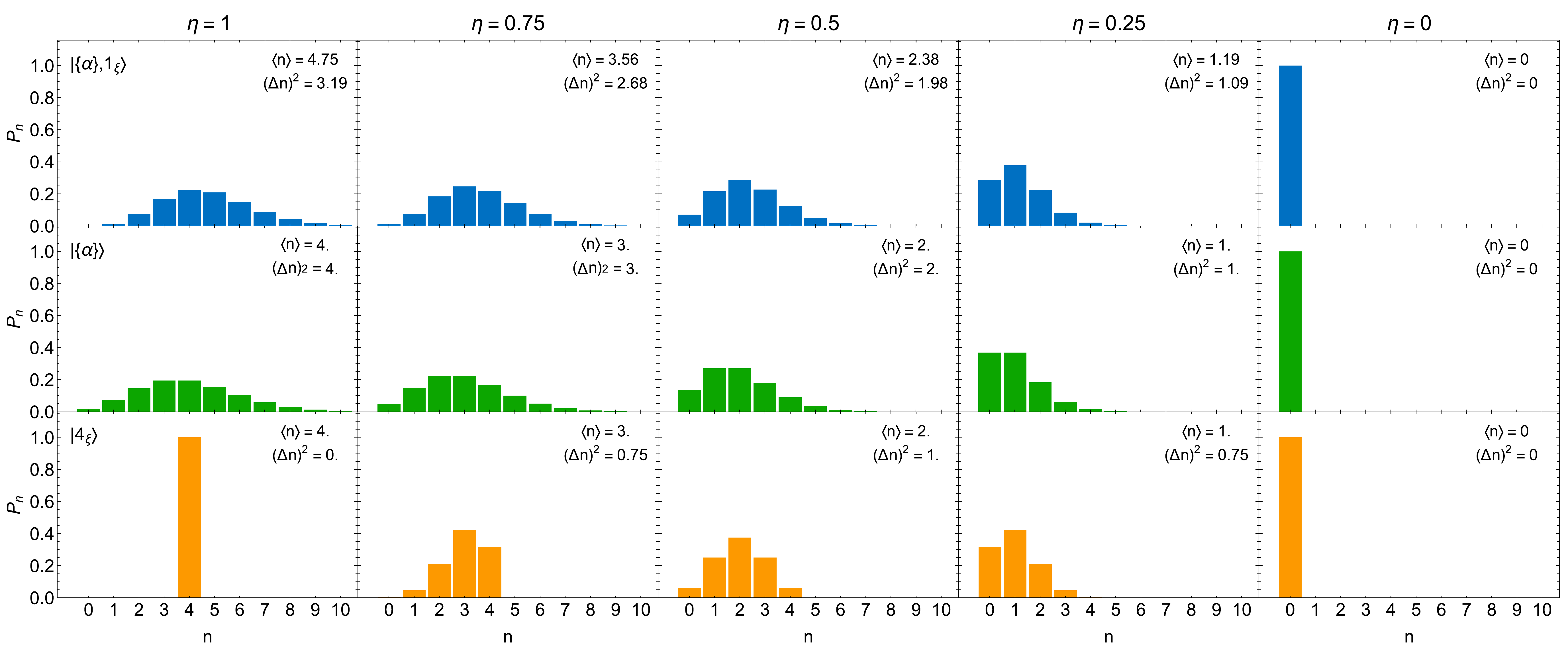}
\caption{Photon number probability distributions for a continuous-mode PAC state as it propagates in a waveguide and undergoes loss. Top row: a PAC state with $n_\alpha=3$, middle row: a coherent state with $n_\alpha=4$, and bottom row: a Fock state with $n=4$. The loss increases from left to right and is shown above each column. The corresponding propagation distances, $L$, for a plasmonic nanowire (n), stripe (s), and dielectric fibre (d), as shown in Fig.~\ref{fig1}~(c), are $L_{\rm n}=L_{\rm s}=L_{\rm d}=0$ ($\eta=1$); $L_{\rm n}=0.35\mu$m, $L_{\rm s}=4.32\mu$m and $L_{\rm d}=2.88$km ($\eta=0.75$); $L_{\rm n}=0.83\mu$m, $L_{\rm s}=10.40\mu$m and $L_{\rm d}=6.93$km ($\eta=0.5$);  $L_{\rm n}=1.66\mu$m, $L_{\rm s}=20.79\mu$m and $L_{\rm d}=13.86$km ($\eta=0.25$);  $L_{\rm n}=L_{\rm s}=L_{\rm d}=\infty$ ($\eta=0$). The mean photon number and variance are also shown for each distribution as insets.}
\label{fig2} 
\end{figure*}

We start by calculating the photon number probability distribution, mean, and variance for PAC states. As outlined earlier, we begin with the probability density,
	\begin{equation}
		P_n\big(\{t_i\}_{i=1}^{n}\big)=\frac{|N(\tau)|}{n!}|\braket{0|\prod_{i=1}^{n}\hat{a}(t_i)\hat{a}_\xi^\dagger|\{\alpha\}}|^2,
	\end{equation}
which has been expanded from Eq.~\eqref{eqn:Pdensity} using the state definitions in Eqs.~\eqref{eqn:nt} and \eqref{eqn:pacs}. Writing the operator product in normal order using Eq.~\eqref{eqn:atnaxi} and then carrying out the operations we obtain,
\begin{equation}
\begin{split}
	P_n\big(\{t_i\}_{i=1}^{n}\big) =& \frac{|N(\tau)|}{n!}\Big|\sum_{k=1}^{n}\Big[\xi(t_k+\tau)\prod_{\stackrel{i=1}{i\neq k}}^{n}\alpha(t_i)\Big]\Big|^2\mathrm{e}^{-n_\alpha}\\
	= &\frac{|N(\tau)|}{n!}\Bigg(\sum_{k=1}^{n}{|\xi(t_k+\tau)|}^2~\prod_{\substack{i=1\\i\neq k}}^{n}{|\alpha(t_i)|}^2 ~+\\& ~\sum_{k=1}^{n}\sum_{\substack{k'=1\\k'\neq k}}^{n}~\xi(t_k+\tau)~\alpha^*(t_k)~\xi^*(t_{k'}+\tau)~\alpha(t_{k'})\\&
	\times\prod_{\substack{i=1\\i\neq k,k'}}^{n}~{|\alpha(t_i)|}^2\Bigg)\mathrm{e}^{-n_\alpha}.
\end{split}
\label{eqn:Pdensity_pacs}
\end{equation}
The modulus-squared factor has been separated into squared and mixed terms. In this form, the integral of Eq.~\eqref{eqn:Pdensity_pacs} with respect to all times simplifies easily. Performing the integration gives the photon number probability distribution,
\begin{equation}
\begin{split}
P_n =&~\frac{|N(\tau)|}{n!}\Bigg(\sum_{k=1}^{n}\int dt_k~{|\xi(t_k+\tau)|}^2~\prod_{\substack{i=1\\i\neq k}}^{n}\int dt_i{|\alpha(t_i)|}^2 ~+\\& ~\sum_{k=1}^{n}\sum_{\substack{k'=1\\k'\neq k}}^{n}~\int dt_k~\xi(t_k+\tau)~\alpha^*(t_k)\int dt_{k'}~\xi^*(t_{k'}+\tau)~\alpha(t_{k'})\\&
\times\prod_{\substack{i=1\\i\neq k,k'}}^{n}~\int dt_i~{|\alpha(t_i)|}^2\Bigg)\mathrm{e}^{-n_\alpha}.
\end{split}
\end{equation}
Substituting Eqs.~\eqref{eqn:modsqdxi}, \eqref{eqn:modsqdalpha}, \eqref{eqn:sigma} and its complex conjugate for the integrals, and simplifying we obtain
\begin{equation}
	P_n = |N(\tau)|~\frac{n_\alpha^{n-1}}{(n-1)!}~\mathrm{e}^{-n_\alpha}~\Bigg(1 + (n - 1)~\frac{{|\sigma(\tau)|}^2}{n_\alpha}\Bigg).
	\label{eqn:Pn_pacs}
\end{equation}
In the limit of perfect temporal overlap, {\it i.e.} $\tau=0$ and $\Omega_1=\Omega$, we recover the single-mode result given in Ref.~\cite{Zavatta05}. 

To obtain an expression for the distribution after propagation, we substitute Eq.~\eqref{eqn:Pn_pacs} into Eq.~\eqref{eqn:PnL} and find that the series converges, yielding
\begin{equation}
\begin{split}
P_n(L) = &~{|\eta(L)|}^n~|1 - \eta(L)|~|N(\tau)|~\mathrm{e}^{-|\eta(L)|n_\alpha}\frac{{n_\alpha}^n}{n!}~\\
&\times\Big(1 + \frac{2~n~|\sigma(\tau)|^2}{n_\alpha} + |1 - \eta(L)|~|\sigma(\tau)|^2\Big)\\
&+{|\eta(L)|}^n~|N(\tau)|~\mathrm{e}^{-|\eta(L)|n_\alpha}\frac{{n_\alpha}^{n-1}}{(n-1)!}  \\
&\times \Big(1 + (n-1)\frac{|\sigma(\tau)|^2}{n_\alpha} \Big).
\end{split}
\label{eqn:PnL_pacs}
\end{equation}
\begin{figure}[t]
\centering
\includegraphics[width=8.5cm]{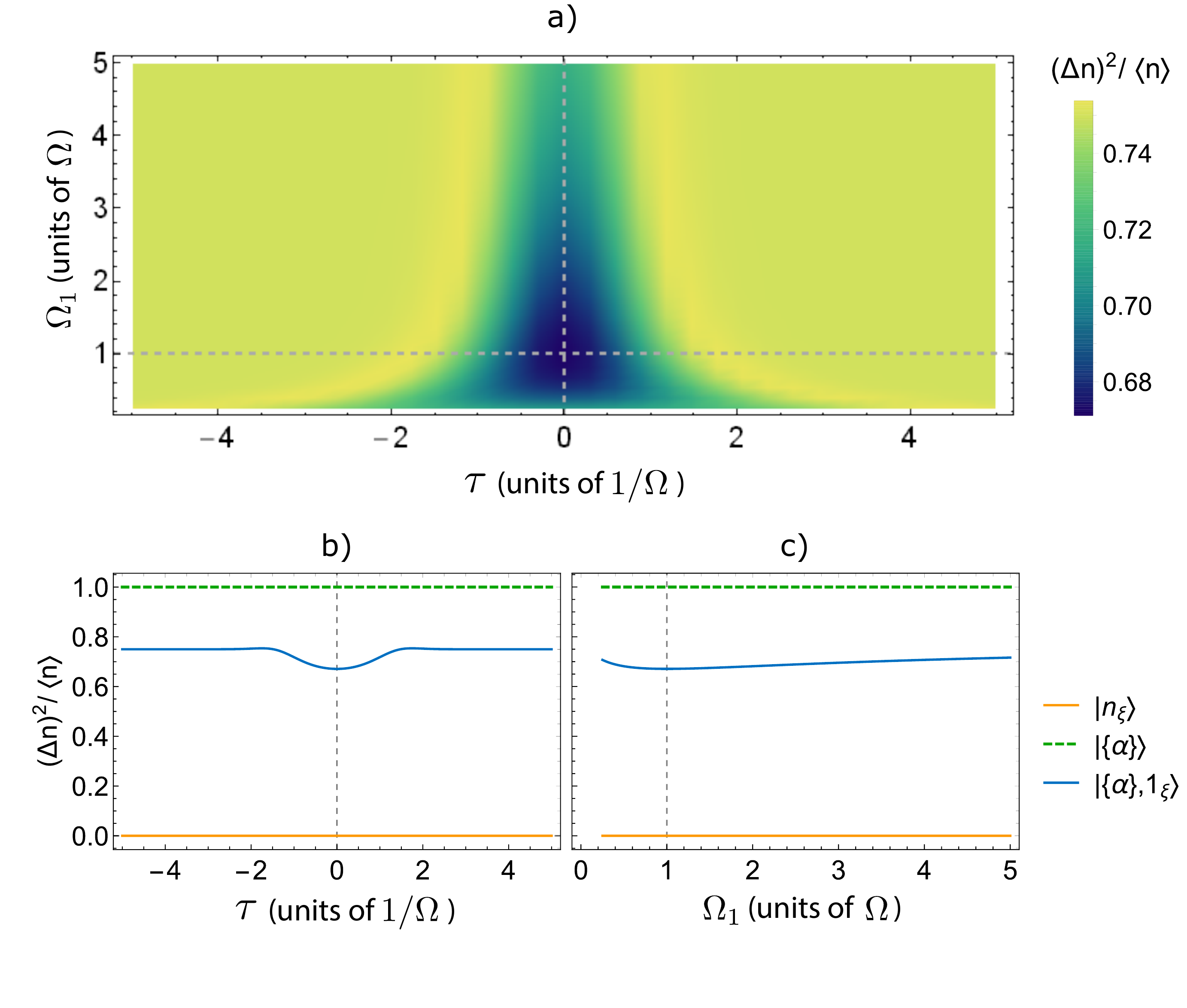}
\caption{Sub-Poissonian behaviour of continuous-mode PAC states using the ratio of the variance to mean of the number state distribution, $(\Delta n)^2/\langle n\rangle$. The added single-photon pulse has a temporal offset $\tau$ in units of $1/\Omega$ and bandwidth $\Omega_1$ in units of $\Omega$, where $\Omega$ is the bandwidth of the coherent state pulse. The central frequency $\omega_0$ has no impact on any expressions plotted, however its value should be at least $10\Omega$ in order to satisfy the narrowband approximation. (a) A density plot showing the variance to mean ratio as $\tau$ and $\Omega_1$ are varied. In (b) and (c), cross sections of the plot in (a) are shown for the PAC state (solid, blue). Also included in these cross sections are the ratios for the number state (solid, orange) and coherent state (dashed, green). The number state and coherent state ratios are independent from $\tau$, $\Omega_1$, and $n_\alpha$, but are shown as a reference. The parameters not being varied in these plots are $n_\alpha=3$ and $|\eta|=1$ (no loss).}
\label{fig3} 
\end{figure}

In Fig.~\ref{fig2} we show histograms of the photon number distribution for a PAC state (first row) with $\tau=0$, $\Omega_1 = \Omega$ and $n_\alpha=3$. The value of $n_\alpha$ is chosen as an example due to it giving the greatest amount of quadrature squeezing in the single-mode treatment~\cite{Zavatta04}. The histograms for various values of $|\eta(L)|$ clearly show that, with increasing loss, the distribution shifts and skews toward lower photon numbers. A coherent state with $n_\alpha=4$ and a four-photon Fock state are also shown for comparison in the middle and bottom rows, respectively. The analytical forms of the probability distributions for these continuous-mode states are given in Appendix \ref{appendix:P}. 

The histograms highlight that the PAC state has an increased mean and a narrower distribution (variance) initially compared to the coherent state with the same mean photon number and is therefore sub-Poissonian~\cite{Loudon00}. One can clearly see how the mean and variance of the PAC state are affected by loss compared to the other states. The figure illustrates the basic number statistics of the continuous mode PAC state when there is perfect temporal and bandwidth overlap.

We now study the effect of imperfect temporal overlap, focussing on the sub-Poissonian nature of PAC states, which is not clearly demonstrated by photon number distribution histograms. Thus, we extract the mean of the distribution in Eq.~\eqref{eqn:PnL_pacs} using $\sum_{n=0}^{\infty}nP_n(L)$, which converges to the result
\begin{equation}
	\langle n \rangle = |\eta(L)|\Big(1 + n_\alpha + {|\sigma(\tau)|}^2~|N(\tau)|\Big).
	\label{eqn:Pn_mean}
\end{equation}
Similarly, using $\sum_{n=0}^{\infty}(n - \braket{n})^2P_n(L)$, the variance is given by
\begin{equation}
	\begin{split}
	{(\Delta n)}^2 = &~\langle n \rangle - \Big(1 - {|\sigma(\tau)|}^4~{|N(\tau)|}^2\Big){|\eta(L)|}^2.
	\end{split}
	\label{eqn:Pn_variance}
\end{equation}
Again, in the limit of perfect temporal overlap and no loss, it is straightforward to verify that the above expressions give the single-mode results for the mean and variance of the photon number distribution~\cite{Agarwal91}. With perfect overlap and in the limit of large $n_\alpha$, ${|\sigma(\tau)|}^4~{|N(\tau)|}^2 \rightarrow 1$. Thus for a coherent state with a large mean photon number, the variance is negligibly close to the mean. The second term, which reduces the variance, is quadratic in $|\eta(L)|$, while the mean is linear. Therefore, as loss increases ($|\eta(L)|$ decreases) the reduction of the variance decays faster than the mean, and the variance to mean ratio increases linearly. 

In Fig.~\ref{fig3}~(a), the ratio of the variance and mean are plotted as the temporal offset $\tau$ and bandwidth $\Omega_1$ of the added single photon in the PAC state are varied (with $n_\alpha=3$). The loss is set to zero for the moment. In the cross section cuts in Figs.~\ref{fig3}~(b) and (c) we have included the results for the coherent state with arbitrary mean photon number, which is Poissonian (variance equal to the mean) and serves as an upper-bound for sub-Poissonian states (variance less than the mean). Also included is an $n$-photon state which, having a variance of $(\Delta n)^2 = 0$, is an ideal sub-Poissonian state. As can be seen in Fig.~\ref{fig3}~(a), the ratio $ (\Delta n)^2/\langle n \rangle < 1$ for all parameter ranges, thus demonstrating the sub-Poissonian nature of PAC states. 

In Fig.~\ref{fig3}~(b), we see that the PAC state is always sub-Poissonian regardless of the temporal overlap. However, the variance (and ratio) decreases with better overlap, with the lowest variance and ratio at $\tau = 0$. The fact that the ratio is always less than one regardless of the temporal offset of the single photon is due to the populations being derived from integrated values. For example, adding a photon completely out of time (or even incoherently) with the coherent state leads to the $P_n$ distribution shifting up by 1 ($n \to n+1$), which increases the mean, but leaves the variance unchanged. This results in a variance to mean ratio of $n_\alpha/(n_\alpha+1)$, which is always less than 1 and gives the asymptotic value of the ratio. In Fig.~\ref{fig3}~(b) this is given by $3/4=0.75$ for large $\tau$. 

In Fig.~\ref{fig3}~(c), the PAC state is again always sub-Poissonian regardless of the bandwidth mismatch, with a minimum value at $\Omega_1=\Omega$, corresponding to perfect spectral overlap.

It is therefore clear that a sub-Poissonian behavior is not enough on its own to completely determine the quantum, or nonclassical, character of a continuous-mode PAC state and further analysis of its properties is needed.

\subsection{Second-order correlation function}
The second property of continuous-mode PAC states we investigate is the second-order correlation function, $g^{(2)}$, which allows us to go beyond basic photon number statistics and study temporal correlations in the statistics of a state. In particular, the correlation between the intensity of a field at time $t_1$ and at time $t_2$, for a fixed position. A value of less than unity at zero delay ($t_1=t_2$) is only possible for a nonclassical state and therefore we can use it to determine nonclassicality~\cite{Loudon00}. The second-order correlation function is given by~\cite{Loudon00,Gardiner00}
\begin{figure*}[t]
\centering
\includegraphics[width=17.5cm]{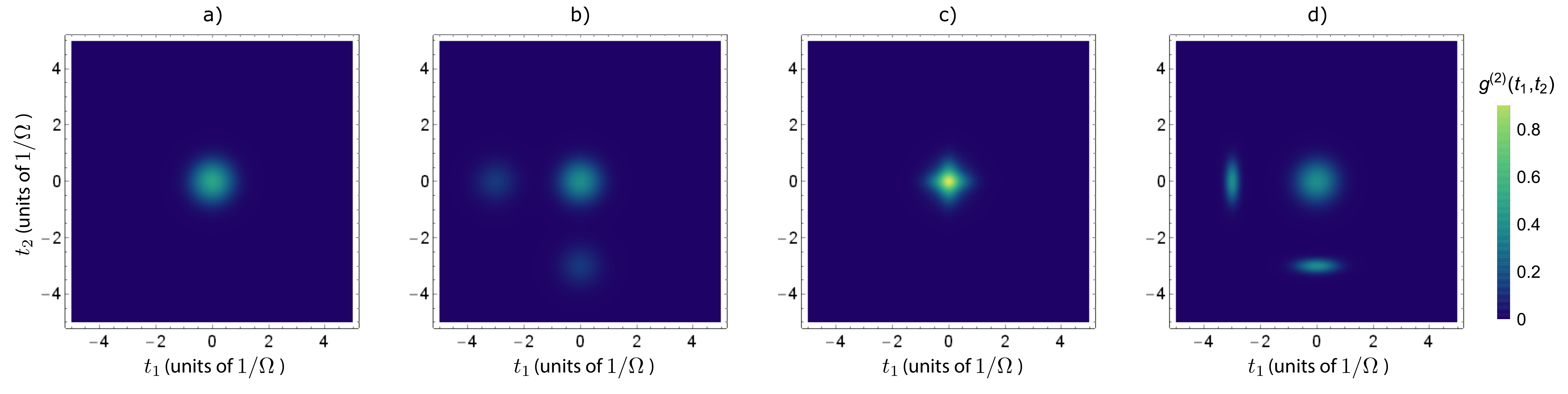}
\caption{Second-order correlation function $g^{(2)}(t_1,t_2)$ of the continuous-mode PAC state for various cases of temporal overlap between the added photon and coherent state. (a) Perfect temporal overlap between the photon and coherent state. (b) The bandwidth of the single photon is the same as the coherent state ($\Omega_1=\Omega$), but the single-photon pulse is shifted in time by $\tau=3/\Omega$. (c) The single photon and coherent state have zero delay, but the bandwidth of the single photon is larger than that of the coherent state (pulse duration shorter) by a factor of three. (d) The bandwidth of the single photon is three times larger than that of the coherent state and its pulse is shifted in time by $\tau=3/\Omega$. 
In all cases the second-order correlation function has a value at zero time delay ($t_1=t_2$) of less than 1, confirming nonclassicality of the state.}
\label{fig5} 
\end{figure*}

\begin{equation}
g^{(2)}(t_1,t_2) = \frac{\braket{\hat{a}^\dagger(t_1)\hat{a}^\dagger(t_2)\hat{a}(t_2)\hat{a}(t_1)}}{\braket{\hat{a}^\dagger(t_1)\hat{a}(t_1)}\braket{\hat{a}^\dagger(t_2)\hat{a}(t_2)}}.
\label{eqn:g2t}
\end{equation}
In this form, $g^{(2)}(t_1,t_2)$ represents the average measure of correlation of the second order of the field at the set of times $\{ t_1,t_2\}$. However, when pulses are involved, as in an experiment, it is also useful to consider a `measured' second-order correlation function at zero time delay, given by~\cite{Loudon00}
\begin{equation}
g^{(2)}[0] = \frac{\int dt_1 dt_2\braket{\hat{a}^\dagger(t_1)\hat{a}^\dagger(t_2)\hat{a}(t_2)\hat{a}(t_1)}}{\int dt_1\braket{\hat{a}^\dagger(t_1)\hat{a}(t_1)}\int dt_2\braket{\hat{a}^\dagger(t_2)\hat{a}(t_2)}}.
\label{eqn:g20}
\end{equation}
The measured version can be obtained from the detection events in an experiment, where detectors monitor photocounts over a fixed period of time, $T$, which encompasses the pulse duration. We will consider both $g^{(2)}(t_1,t_2)$ and $g^{(2)}[0]$. Further details about these two types of second-order correlation and how they are related can be found in Ref.~\cite{Fischer16}.

We start by identifying the constituent quantities of Eqs.~\eqref{eqn:g2t} and \eqref{eqn:g20}, for the case of no loss, as the photon flux,
\begin{equation}
f_1(t) = \braket{\hat{a}^\dagger(t)\hat{a}(t)},
\end{equation}
and two-time coincidence rate,
\begin{equation}
f_2(t_1, t_2) = \braket{\hat{a}^\dagger(t_1)\hat{a}^\dagger(t_2)\hat{a}(t_2)\hat{a}(t_1)}.
\end{equation}
Following the same procedure as was done for the photon number distribution, we arrive at the PAC state photon flux,
\begin{equation}
\begin{split}
f_1(t) =&~ |N(\tau)|\braket{\{\alpha\}|\hat{a}_\xi\hat{a}^\dagger(t)\hat{a}(t)\hat{a}_\xi^\dagger|\{\alpha\}}\\
=&~~|\alpha(t)|^2 + 2|N(\tau)||\sigma(\tau)||\xi(t+\tau)||\alpha(t)| \\ & \qquad + |N(\tau)||\xi(t+\tau)|^2,
\end{split}
\label{eqn:f1t_pacs}
\end{equation}
and coincidence rate,
\begin{equation}
\begin{split}
f_2(t_1,t_2) &=~ |N(\tau)|\Big[|\xi(t_1+\tau)|^2|\alpha(t_2)|^2 + |\xi(t_2+\tau)|^2|\alpha(t_1)|^2\\ 
&+~ |\alpha(t_1)|^2|\alpha(t_2)|^2/|N(\tau)|\\
&+~ 2|\xi(t_1 + \tau)||\xi(t_2 + \tau)||\alpha(t_1)||\alpha(t_2)|\\
&+~ 2|\sigma(\tau)||\xi(t_1 + \tau)||\alpha(t_1)||\alpha(t_2)|^2\\
&+~ 2|\sigma(\tau)||\xi(t_2 + \tau)||\alpha(t_2)||\alpha(t_1)|^2\Big].
\end{split}
\label{eqn:f2t_pacs}
\end{equation}
The second-order correlation function, $g^{(2)}(t_1,t_2)$, is then
\begin{equation}
g^{(2)}(t_1, t_2) = \frac{f_2(t_1, t_2)}{f_1(t_1) f_1(t_2)}.
\label{eqn:g2t_pacs}
\end{equation}
To include the effects of propagation and loss the operators are transformed, as carried out in the previous section, using $\hat{a}(t)\to\hat{a}_L(t)$. We then have that $f_1(t_i) \to |\eta (L)| f_1(t_{r,i})$ and $f_2(t_1,t_2) \to |\eta (L)|^2 f_2(t_{r,1},t_{r,2})$, where $t_{r,i}=t_i - L/v_g$ is a retarded time. Thus, $g^{(2)}(t_1,t_2)$ is unaffected by loss as the loss factors cancel in the numerator and denominator, and the resulting propagation only shifts the time arguments.

In Fig.~\ref{fig5} we show $g^{(2)}(t_1,t_2)$ for various cases of temporal overlap between the added photon and coherent state, with $n_\alpha=3$. Fig.~\ref{fig5}~(a) shows the case of perfect temporal overlap between the photon and coherent state. The correlation function is rotationally symmetric in time due to the wavepackets of the single-photon and coherent state being rotationally symmetric and having perfect overlap. Fig.~\ref{fig5}~(b) shows the case where the bandwidth of the single photon is the same as the coherent state ($\Omega_1=\Omega$), but the single-photon pulse is shifted in time by $\tau=3/\Omega$. The function is no longer rotationally symmetric due to a time offset in the wavepackets. Fig.~\ref{fig5}~(c) shows the case where the single photon and coherent state have zero time delay, but the bandwidth of the single photon is three times larger than the coherent state (shorter duration). Finally, Fig.~\ref{fig5}~(d) shows the case where the bandwidth of the single photon is three times larger than that of the coherent state and its pulse is shifted in time by $\tau=3/\Omega$. 

In all cases, the second-order correlation function has a value at zero time delay ($t_1=t_2$, {\it i.e.} along the diagonal) of less than 1, confirming non-classicality of the state. This is in contrast to a coherent state pulse, which gives $g^{(2)}(t_1,t_2)=1~\forall~t_1,t_2$, and a single-photon pulse, which gives $g^{(2)}(t_1,t_2)=0 ~\forall ~t_1,t_2$~\cite{Loudon00}. The fact that $g^{(2)}(t_1,t_2)$ is constant over all time for these states, even though they are represented by a Gaussian pulse is due to $g^{(2)}(t_1,t_2)$ being a ratio of a coincidence rate and a photon flux of a wavepacket state that theoretically extends over all time, with the amplitudes of the coincidence rate and flux cancelling to give a constant at all times. On the other hand, $g^{(2)}(t_1,t_2)$ for the PAC state is not uniform like the coherent and single-photon states, but has localised hotspots, corresponding to pairs of times where there is more likely to be a coincidence. These additional coincidences are those between the added single-photon and photons of the coherent state pulse. Thus, the location and shape of the hotspots are determined by the temporal offset and width of the single photon with respect to the coherent state. The value of $g^{(2)}(t_1,t_2)$ remains approximately zero outside these hotspots due to the relative amplitudes of the coincidence rate and flux.

To calculate the measured version of the second-order correlation, $g^{(2)}[0]$, we must perform the integrations. For no loss, integrating Eq.~\eqref{eqn:f1t_pacs} gives the mean photon number 
\begin{equation}
f_1 = 1 + n_\alpha + |N(\tau)||\sigma(\tau)|^2,
\label{eqn:f1_pacs}
\end{equation}
which matches Eq.~(\ref{eqn:Pn_mean}) for $|\eta(L)|=1$. Integrating Eq.~\eqref{eqn:f2t_pacs} over both times gives the average number of coincidences
\begin{equation}
\begin{split}
f_2 = n_\alpha^2 + 2n_\alpha + 2|N(\tau)||\sigma(\tau)|^2\big(n_\alpha + 1\big).
\end{split}
\label{eqn:f2_pacs}
\end{equation}
Taking the ratio $f_2/f_1^2$ then gives
\begin{equation}
\begin{split}
	g^{(2)}[0] = & ~\frac{n_\alpha^2 + 2n_\alpha + 2|N(\tau)||\sigma(\tau)|^2\big(n_\alpha + 1\big)}{\Big(1 + n_\alpha + |N(\tau)||\sigma(\tau)|^2\Big)^2}\\
	= & ~1 - \frac{1 + |N(\tau)|^2|\sigma(\tau)|^4}{\Big(1 + n_\alpha + |N(\tau)||\sigma(\tau)|^2\Big)^2}.
\end{split}
\label{eqn:g20_pacs}
\end{equation}
When loss is included, the loss factors cancel for $f_1^2$ and $f_2$, as in the case of $g^{(2)}(t_1,t_2)$. Note that in this case there is no retarded time as all times have been integrated. From Eq.~\eqref{eqn:g20_pacs}, it is clear that $g^{(2)}[0] < 1$ for all parameters as $|N(\tau)| > 0$ and $|\sigma(\tau)| \geq 0$ always. Furthermore, the second term is always less than $1$ for non-zero $n_\alpha$. In the limit of large $n_\alpha$, the second term approaches zero resulting in a $g^{(2)}[0]$ value approaching $1$, as would be expected with a strong coherent state. 

\begin{figure}[t]
	\centering
	\includegraphics[width=8.6cm]{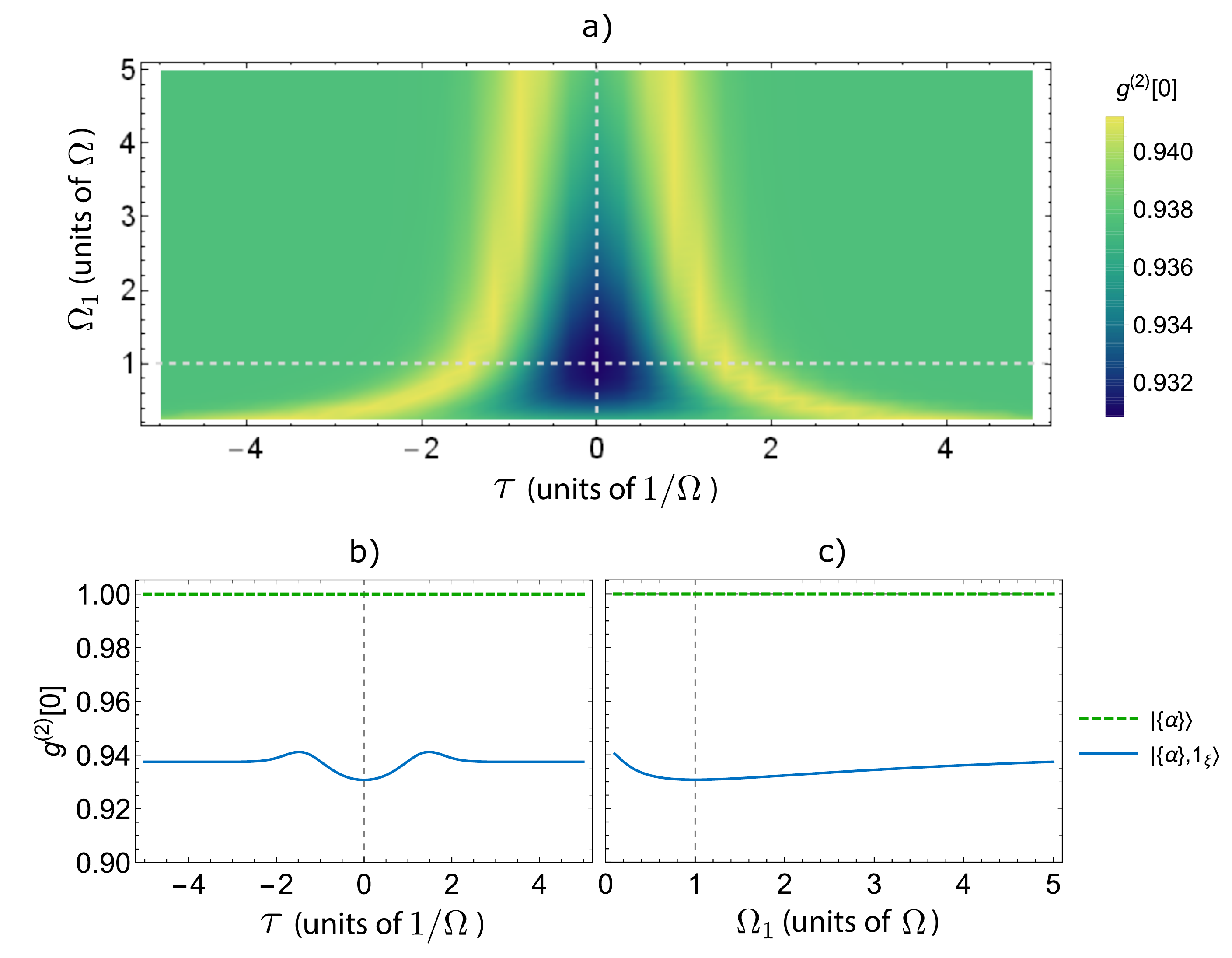}
	\caption{Measured second-order correlation function, $g^{(2)}[0]$, of continuous-mode PAC states. The added single-photon pulse has a temporal offset $\tau$ in units of $1/\Omega$ and bandwidth $\Omega_1$ in units of $\Omega$, where $\Omega$ is the bandwidth of the coherent state pulse. (a) Density plot showing $g^{(2)}[0]$ as $\tau$ and $\Omega_1$ are varied. In (b) and (c), cross sections of the plot in (a) are shown for the PAC state (solid, blue). Also included in these cross sections are $g^{(2)}[0]$ values for the coherent state (dashed, green). The values are independent from $\tau$, $\Omega_1$, and $n_\alpha$, but are shown as a reference. The value of the mean photon number for the coherent state in the PAC state is $n_\alpha=3$.}
	\label{fig6} 
\end{figure}

In Fig.~\ref{fig6}~(a) we show $g^{(2)}[0]$ as the temporal offset, $\tau$, and bandwidth, $\Omega_1$, are varied. The mean photon number of the coherent state within the PAC state is  $n_\alpha=3$ as an example. In Figs.~\ref{fig6}~(b) and (c) we show cross sections corresponding to $\Omega_1=\Omega$ and $\tau=0$, respectively. Also shown are values of $g^{(2)}[0]$ for a coherent state (arbitrary mean photon number). One can see that the value of $g^{(2)}[0]$ is less than 1 for the PAC state for all values of $\tau$ and $\Omega_1$, confirming its nonclassicality.

The study of the second-order correlation function complements that of the photon number statistics for determining the nonclassical character of a continuous-more PAC state. However, both the photon number statistics (sub-Poissonian behaviour) and the second-order correlation do not completely characterize a given PAC state, even though they highlight its nonclassicality well. 

\section{Quadratures}
We now move on from studying the basic statistics of the field and derive the mean and variance of the quadrature operator. The aim is to determine if, and under what conditions, the continuous-mode PAC state $\ket{\{\alpha\},1_\xi}$ may be quadrature squeezed. Such a state would have one quadrature with a variance less than that of a coherent state quadrature and is another signature of nonclassicality~\cite{Lvovsky15}.

The PAC state is formally described as a non-Gaussian state~\cite{Tan19}, as its Wigner function is not a Gaussian distribution~\cite{Agarwal91}. As such, the mean and variance of the quadrature operator do not completely characterize it, unlike for Gaussian states. However, they represent additional physical quantities beyond the basic sub-Poissonian statistics and second-order correlations that can be used to characterize a state's nonclassicality. In general, Gaussian and non-Gaussian states are important in a wide range of applications in quantum information processing~\cite{Tan19}. In particular, in quantum sensing, where the squeezed nature of a state gives intra-mode correlations that can be exploited to improve precision measurements of phase~\cite{Knott16}. This has been studied for non-Gaussian states~\cite{Braun14}, where it was found that non-Gaussian states generated by Gaussian states (e.g. a coherent state) modified by the subtraction and addition of photons provides enhanced sensitivity in phase sensing. Thus, the squeezed nature of non-Gaussian states such as PAC states can provide advantages under certain conditions and it is therefore an important aspect to study.

The derivation of the mean and variance of the continuous-mode coherent state and number state quadratures are given in Appendix \ref{appendix:Q}. Here, we focus on the PAC state.

We begin with the instantaneous quadrature operator~\cite{Loudon00},
	\begin{equation}
		\hat{X}_\varphi(t) = \frac{1}{2} \Big(~\at \mathrm{e}^{-\im \varphi(t)} + \atd \mathrm{e}^{\im \varphi(t)}~\Big),
		\label{eqn:Xphit}
	\end{equation}
where $\varphi(t)$ is the quadrature phase. Similar to the measured second-order correlation function, this definition is integrated with respect to time, making it suitable in an experimental context where measurements usually span a period of time, $T$. Taking the expectation value of Eq.~\eqref{eqn:Xphit} and then integrating, we obtain the total mean quadrature,
	\begin{equation}
		\braket{\hat{X}_\varphi(t,T)} = \int_{t}^{t + T} dt'~\mathrm{Re}\Big\{\braket{\hat{a}(t')} \mathrm{e}^{-\im\varphi(t')}\Big\}.
		\label{eqn:Xphi_mean}
	\end{equation}
The conjugate quadrature $\hat{X}_{\varphi+\pi/2}(t, T)$ is obtained by substituting $\varphi(t) \rightarrow \varphi(t) + \pi/2$ in Eq.~\eqref{eqn:Xphit}.

The variance of the quadrature operator is given by $\Big(\Delta X_\varphi(t,T)\Big)^2 = \braket{{\hat{X}_\varphi}^2(t,T)} - \braket{\hat{X}_\varphi(t,T)}^2$. This expression becomes simpler to use in the normal-ordered form~\cite{Blow90}:
\begin{equation}
\Big(\Delta X_\varphi(t,T)\Big)^2 = \frac{T}{4} + \braket{:{\hat{X}_\varphi}^2(t,T):} - \braket{\hat{X}_\varphi(t,T)}^2,
\label{eqn:Xphi_var}
\end{equation}
where
\begin{equation}
\begin{split}
	\braket{:{\hat{X}_\varphi}^2(t,T):} = &\frac{1}{2} \int_{t}^{t + T} dt'dt''~\mathrm{Re}\Big\{\braket{\hat{a}(t')\hat{a}(t'')}\mathrm{e}^{-\im \{\varphi(t') + \varphi(t'')\}} \\
	&\qquad+ \braket{\hat{a}^{\dagger}(t'')\hat{a}(t')}\mathrm{e}^{-\im \{\varphi(t') - \varphi(t'')\}} \Big\}.
\end{split}
\label{eqn:Xphi2moment}
\end{equation}
The expression in Eq.~(\ref{eqn:Xphi_var}) can be derived by squaring Eq.~\eqref{eqn:Xphit}, then normal-ordering the result using the commutation relation in Eq.~\eqref{eqn:atcomm} and taking the expectation value to obtain $\braket{{\hat{X}_\varphi}^2(t,T)} = T/4 + \braket{:{\hat{X}_\varphi}^2(t,T):}$.

We now study the quadrature mean and variance with no propagation loss for PAC states using the above formulas. The mean can be calculated by evaluating Eq.~\eqref{eqn:Xphi_mean} with respect to the state $\ket{\{\alpha\},1_\xi}$, which is expanded using Eq.~\eqref{eqn:pacs}, giving
\begin{equation}
	\braket{\hat{X}_\varphi(t,T)} = |N(\tau)|\int_{t}^{t + T}dt'\mathrm{Re}\Big\{\braket{\{\alpha\}|\hat{a}_\xi\hat{a}(t')\hat{a}_\xi^\dagger|\{\alpha\}} \mathrm{e}^{\im \varphi(t')}\Big\}.
	\label{eqn:Xphimean_pacs0}
\end{equation}
The operator product can be re-ordered as, 
\begin{equation}
	\hat{a}_\xi\hat{a}(t')\hat{a}_\xi^\dagger = \xi(t' + \tau)\hat{a}_\xi + \hat{a}(t') + \hat{a}_\xi^\dagger\hat{a}_\xi\hat{a}(t'),
\end{equation}
using Eqs.~\eqref{eqn:axicom} and \eqref{eqn:atnaxi}. Then carrying out these operations on $\ket{\{\alpha\}}$ using Eqs. \eqref{eqn:eigen} and \eqref{eqn:axieig}, we obtain
\begin{equation}
\begin{split}
\bra{\{\alpha \}}\hat{a}_\xi\hat{a}(t')\hat{a}_\xi^\dagger\ket{\{\alpha\}} =\sigma(\tau)\xi(t' + \tau) + \alpha(t')(1 + |\sigma(\tau)|^2)\\
 = \Big(|\sigma(\tau)||\xi(t' + \tau)|+ |\alpha(t')|\big(1 + |\sigma(\tau)|^2\big)\Big)\mathrm{e}^{-\im \theta(t')}.
\end{split}
\end{equation}
Substituting this into Eq.~\eqref{eqn:Xphimean_pacs0} and taking the real part, gives
\begin{equation}
	\begin{split}
	\braket{\hat{X}_\varphi(t, T)} =&~|N(\tau)||\sigma(\tau)|\int_{t}^{t + T}dt'~|\xi(t' + \tau)|\cos\big[\theta(t') - \varphi(t')\big]\\
	& + \int_{t}^{t + T}dt'~|\alpha(t')|\cos\big[\theta(t') - \varphi(t')\big].
	\end{split}
	\label{eqn:Xphimeanpacs}
\end{equation}
In the same manner, $\braket{:\hat{X}_\varphi^2(t, T):}$ can be derived;
\begin{equation}
\begin{split}
	\braket{:\hat{X}_\varphi^2(t, T):} &=~ \braket{\hat{X}_\varphi(t,T)}^2 -{|N(\tau)|}^2{|\sigma(\tau)|}^2 \\
&	\times\Bigg[\int_{t}^{t + T}dt'~|\xi(t' + \tau)|\cos\big[\theta(t') - \varphi(t')\big]\Bigg]^2\\
&	+~\frac{1}{2}|N(\tau)|\int_{t}^{t+T}dt'dt''|\xi(t' + \tau)||\xi(t'' + \tau)|\\
&	\times\cos\big[\theta(t') - \varphi(t') - \theta(t'') + \varphi(t'')\big].
\end{split}
\label{eqn:Xphi2moment_pacs}
\end{equation}
By inserting Eqs.~\eqref{eqn:Xphimeanpacs} and \eqref{eqn:Xphi2moment_pacs} into Eq.~\eqref{eqn:Xphi_var}, we obtain 
\begin{equation}
	\begin{split}
		&\Big(\Delta X_\varphi(t, T)\Big)^2  = ~\frac{T}{4} 
		+ |N(\tau)|\Big(\frac{1}{2} - |N(\tau)||\sigma(\tau)|^2\Big)\\
		&\qquad \qquad\times\Bigg[\int_{t}^{t + T}dt'~|\xi(t'+\tau)|\cos\big[\theta(t') - \varphi(t')\big]\Bigg]^2\\
		&\qquad \qquad+\frac{1}{2}|N(\tau)|\Bigg[\int_{t}^{t + T}dt'~|\xi(t'+\tau)|\sin\big[\theta(t') - \varphi(t')\big]\Bigg]^2.\\
	\end{split}
	\label{eqn:Xphivar_pacs}
\end{equation}
When Eq.~\eqref{eqn:Xphivar_pacs} is less than the quadrature variance of the coherent state, $T/4$ (see Appendix~\ref{appendix:qCS}), the PAC state can be said to be squeezed. To determine the quadrature most likely to be squeezed, we look for the quadrature phase $\varphi(t)$ such that the variance $(\Delta X_\varphi(t,T))^2$ is minimal. Since only the second term in Eq.~\eqref{eqn:Xphivar_pacs} may be negative, we choose $\varphi(t) = \theta(t)$ to maximise the integral factor while reducing the third term to zero. This gives the average
\begin{equation}
	\braket{\hat{X}_\theta(t, T)} = \int_{t}^{t + T} dt'~\Big(|N(\tau)||\sigma(\tau)||\xi(t' + \tau)| + |\alpha(t')|\Big),
	\label{eqn:Xphiminmean_pacs}
\end{equation}
with the minimal variance
\begin{equation}
\begin{split}
\Big(\Delta X_\theta(t, T)\Big)^2 = \frac{T}{4} + |N(\tau)|\Big(\frac{1}{2} - |N(\tau)|{|\sigma(\tau)|}^2\Big)\\
\times~\Bigg[\int_{t}^{t + T} dt' |\xi(t'+\tau)|\Bigg]^2.
\end{split}
\label{eqn:Xphiminvar_pacs}
\end{equation}
A reduction in the variance is determined by the value of $|N(\tau)||\sigma(\tau)|^2$, with larger values providing higher reduction. Values greater than $1/2$ yield a variance $(\Delta X_\theta(t,T))^2 < T/4$, indicating that the state exhibits quadrature squeezing. In the case of perfect overlap $|N(\tau)||\sigma(\tau)|^2 = n_\alpha/(1 + n_\alpha)$, which is greater than $1/2$ for $n_\alpha > 1$, with the corresponding variance expression resembling the single-mode result~\cite{Agarwal91}.

Propagation loss is incorporated into the mean by replacing $\hat{a}(t')$ in Eq.~\eqref{eqn:Xphi_mean} with $\hat{a}_L(t')$ . Then, using Eq.~\eqref{eqn:aL} we obtain 
\begin{equation}
	\begin{split}
		\braket{\hat{X}_\varphi(t,T,L)} &= \int_{t}^{t + T} dt'~\mathrm{Re}\Big\{\Big\langle\eta^{\frac{1}{2}}(L)\hat{a}(t_r')\\
		&~~~~~~~~~~~~~~~~~~~~~+ \im(1-\eta(L))^{\frac{1}{2}}\hat{v}(t')\Big\rangle\mathrm{e}^{-\im\varphi(t')}\Big\},
	\end{split}
\end{equation}
where $t'_r = t' - L/v_g(\omega_0)$. Since the environment is initially in a vacuum state, $\braket{\hat{v}(t')} = 0$, the above expression reduces to
\begin{equation}
	\begin{split}
		\braket{\hat{X}_\varphi(t,T,L)} &=  |\eta(L)|^{1/2} \int_{t + L/v_g}^{t + L/v_g + T} dt'~\mathrm{Re}\Big\{\braket{\hat{a}(t_r')} \mathrm{e}^{-\im\varphi_r(t_r')}\Big\}\\
		&=~|\eta(L)|^{1/2} \braket{\hat{X}_{\varphi_r}(t,T)}.
	\end{split}
	\label{eqn:XphiL_mean}
\end{equation}
Note that we have shifted the integral limits by the propagation time $L/v_g$ to remain centred on the pulse. This allows the mean quadrature with loss to equal the attenuated and phase-shifted lossless quadrature. The phase is shifted, resulting in a retarded phase of $\varphi_r(t'_r)=\varphi(t'_r+L/v_g(\omega_0))-\varphi_\eta/2$.

Applying the same procedure to Eq.~\eqref{eqn:Xphi2moment} as was performed for the mean, gives
\begin{equation}
	\braket{:{\hat{X}_\varphi}^2(t,T,L):} =  ~|\eta(L)|\braket{:{\hat{X}_{\varphi_r}}^2(t,T):}.
	\label{eqn:Xphi2Lmoment}
\end{equation}
Combining this result with the mean in Eq.~\eqref{eqn:XphiL_mean} gives the variance with loss, which can be expressed in terms of the phase-shifted lossless variance as~\cite{Blow90,Loudon00}
\begin{equation}
	\Big(\Delta X_\varphi(t, T, L)\Big)^2 = \frac{T}{4}\big(1 - |\eta(L)|\big) + |\eta(L)|\Big(\Delta X_{\varphi_r}(t, T)\Big)^2.
	\label{eqn:XphiL_var}
\end{equation}

Minimising the lossy variance in Eq.~\eqref{eqn:XphiL_var}, requires that we minimise the phase-shifted lossless variance $(\Delta X_{\varphi_r}(t,T))^2$. Thus, as before, we simply set $\varphi_r(t) = \theta(t)$ in Eqs.~\eqref{eqn:XphiL_mean} and \eqref{eqn:XphiL_var}. This means that the quadrature phase, and that of the local oscillator in a homodyne detection scheme, is $\varphi(t) = \theta(t_r) + \varphi_\eta/2$. The lossy quadrature mean is then
\begin{equation}
	\braket{\hat{X}_\varphi(t,T,L)} = |\eta(L)|^{1/2}\braket{\hat{X}_\theta(t,T)},
	\label{eqn:XphiLmean_pacs}
\end{equation}
with the minimal variance
\begin{equation}
\Big(\Delta X_\varphi(t, T, L)\Big)^2 = \frac{T}{4} + |\eta(L)|\Big[\Big(\Delta X_{\theta}(t, T)\Big)^2-\frac{T}{4}\Big].
\label{eqn:XphiLvar_pacs}
\end{equation}
In Eq.~\eqref{eqn:XphiLvar_pacs} we see that with increasing loss ($|\eta(L)|\rightarrow0^+$) the variance linearly approaches $T/4$ as the initial state goes to the vacuum. Thus, as loss increases, the quadrature variance of a squeezed PAC state linearly increases as the second term in Eq.~\eqref{eqn:XphiLvar_pacs} becomes decreasingly negative.

\begin{figure}[t]
	\centering
	\includegraphics[width=8.6cm]{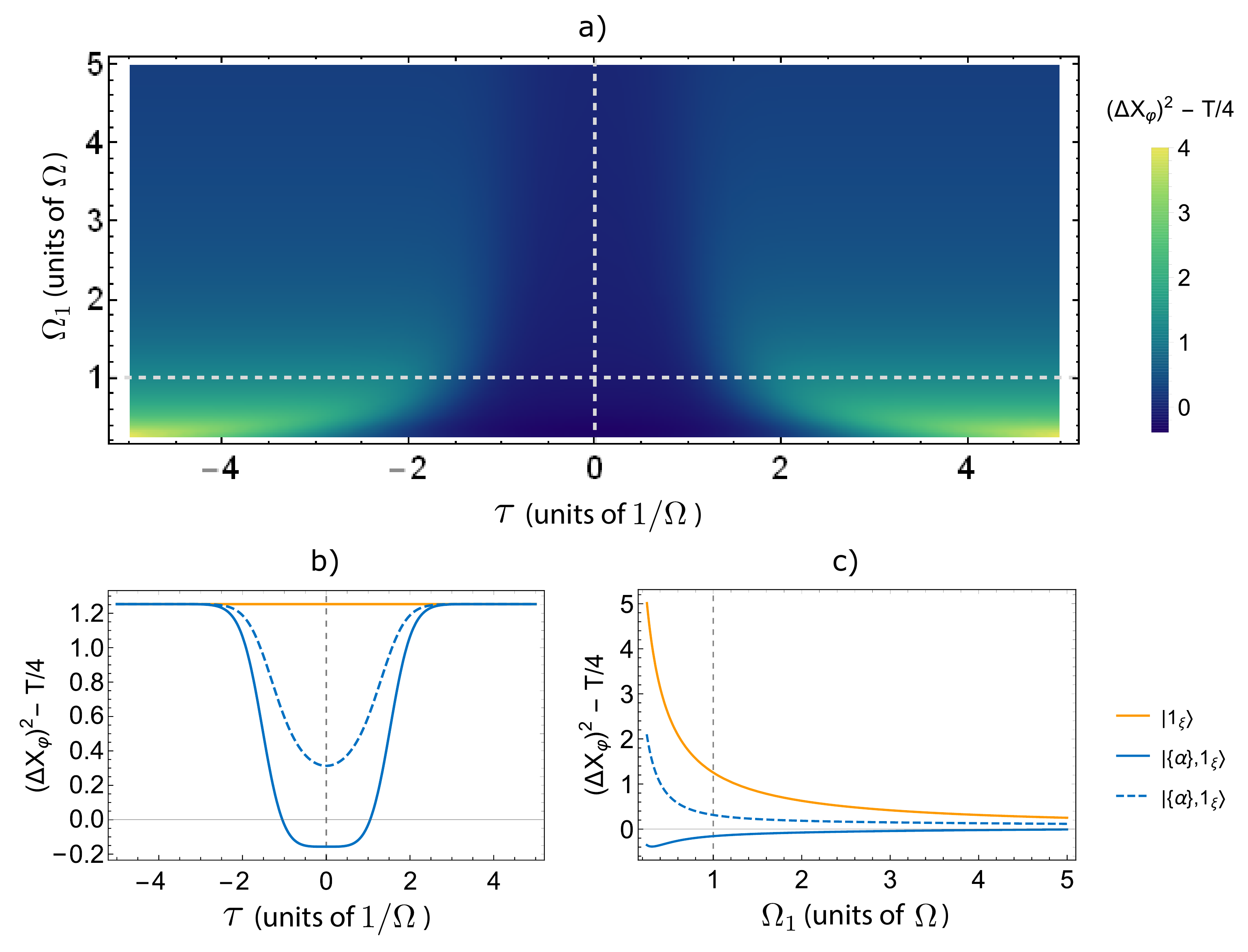}
	\caption{Variance of the quadrature operator for continuous-mode PAC states. (a) The difference in the value of the quadrature variance ($\varphi=\theta$) for continuous-mode coherent states and PAC states as the temporal offset $\tau$ and bandwidth $\Omega_1$ of the added single photon in the PAC state is varied (with $n_\alpha=3$). Negative regions in the plot indicate a parameter regime in which quadrature squeezing occurs. (b) The single-photon offset $\tau$ is varied. (c) The single-photon bandwidth is varied. In (b) and (c) the results for a single-photon state and the out-of-phase quadrature for the PAC state ($\varphi=\theta+\pi/2$) are included as solid orange and dashed blue lines, respectively.}
	\label{fig10} 
\end{figure}

In Fig.~\ref{fig10}~(a) we show the difference in the value of the quadrature variance for continuous-mode coherent states and PAC states given by Eq.~\eqref{eqn:Xphiminvar_pacs} as the temporal offset $\tau$ and bandwidth $\Omega_1$ of the added single photon in the PAC state is varied (with $n_\alpha=3$). The loss is set to zero for the moment. In the cross section cuts in Figs.~\ref{fig10}~(b) and (c) we have included the results for a single-photon state (see Appendix \ref{appendix:qNS}) and the out-of-phase quadrature for the PAC state ($\theta \to \theta+\pi/2$). Negative regions in the plot indicate a parameter regime in which quadrature squeezing occurs. Thus, from Fig.~\ref{fig10}~(b), one can see that the variance of $\hat{X}_\theta$ is less than that of the coherent state, provided that the added single photon is off-set by less than a pulse-width. As the offset increases beyond one pulse-width, the PAC state becomes anti-squeezed, with its variances approaching the number state result. In general, maintaining the indistinguishability of the added single photon from the coherent state photons is required to ensure optimal squeezing. Similarly, in Fig.~\ref{fig10}~(c), as the single-photon bandwidth is broadened it reduces its pulse-width in time, which induces some distinguishability. The PAC state quadrature variance then approaches that of the number state result.

\begin{figure}[t]
	\centering
	\includegraphics[width=6.8cm]{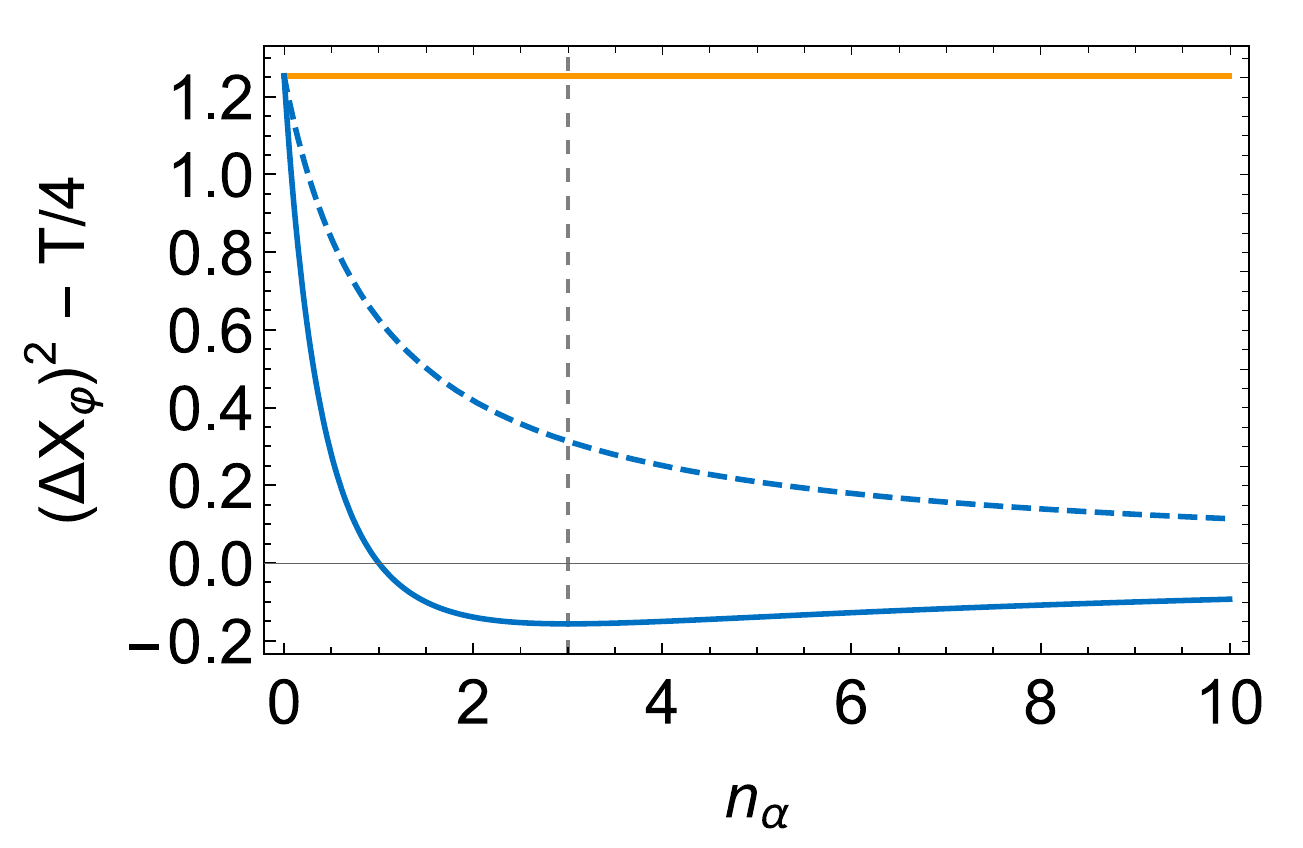}
	\caption{Variance of the quadrature operator ($\varphi=\theta$) for continuous-mode PAC states as the mean photon number of the coherent state, $n_\alpha$, is varied. The dotted vertical line corresponds to $n_\alpha=3$, giving the value obtained in Fig.~\ref{fig10}~(a) at $\tau=0$ and $\Omega_1=\Omega$. Also shown are the single-photon state (solid, orange) and the out-of-phase quadrature for the PAC state (dashed, blue) with $\varphi=\theta+\pi/2$.}
	\label{fig11} 
\end{figure}

In Fig.~\ref{fig11} we consider perfect temporal overlap and show the quadrature variance for continuous-mode PAC states given by Eq.~\eqref{eqn:Xphiminvar_pacs} as the mean photon number of the coherent state, $n_\alpha$, is varied. Also shown are the results for a single-photon state and the out-of-phase quadrature for the PAC state. In Fig.~\ref{fig11}, one can see that as the intensity of the coherent state increases, the less squeezed the PAC state becomes as it begins to resemble a coherent state, a result well known from the single-mode case~\cite{Zavatta04}. Conversely, if the coherent state pulse within the PAC state has fewer than one photon on average, the single-photon pulse dominates resulting in anti-squeezing. The optimal coherent state pulse has $n_\alpha = 3$. By modifying the parameters $\tau$ and $\Omega_1$, behaviour similar to that seen in the plots of Fig.~\ref{fig10}~(b) and (c) are obtained for arbitrary $n_\alpha$, but with overall low values of the quadrature variance for small $n_\alpha$ ($\gtrsim 3$) and higher values for large $n_\alpha$, as well as $n_\alpha < 3$.

When loss is included during propagation, with an initial imperfect pulse overlap, the variance of the quadrature is given by Eq.~\eqref{eqn:XphiLvar_pacs}. Here, changing the pulse parameters simply changes the starting variance at the boundary $|\eta(L)|=1$. As with the case of perfect overlap, the variance linearly increases from its starting value as the second term becomes decreasingly negative for increasing loss.

\section{Fidelity} 

In the previous sections it was found that the sub-Poissonian behaviour, second-order correlation function and quadrature squeezing all highlight the nonclassical nature of continuous-mode PAC states. In the case of sub-Poissonian behaviour and the second-order correlation function, it is interesting to note that for all values of $\tau$ and $\Omega_1$ a PAC state can be said to be nonclassical to a varying degree. The nonclassicality of continuous-mode PAC states is in some sense robust to timing and bandwidth imperfections for certain quantities like these and therefore they do not tell us how good or bad a given state is compared to the ideal case. 

We now introduce a final quantity, the fidelity, with the aim
of characterizing
the `quality' of the PAC state. In this case, the quality is represented by the overlap squared of the PAC state with the ideal case. The fidelity is a measure of the closeness between two states $\ket{\psi}$ and $\ket{\phi}$. It is defined as $F=|\langle \psi | \phi \rangle|^2$~\cite{Nielsen10}, which is equal to 1 for perfect overlap and zero for no overlap. In the case of PAC states we have $\ket{\psi}=(1+n_\alpha)^{-1/2}\hat{a}_{\xi_0}^\dag |\ket{\{ \alpha \}}$ as the ideal PAC state with a single photon pulse profile $\xi_0(t)=\frac{1}{\sqrt{n_\alpha}}\alpha (t)$ having perfect timing and bandwidth, and $\ket{\phi}=\ket{1_\xi,\{ \alpha \}}$ as the PAC state with a single photon profile $\xi$ having arbitrary timing and bandwidth. This leads to
\begin{equation}
F=|\langle \psi | \phi \rangle|^2=\frac{|N(\tau)|}{1+n_\alpha} |\langle \{ \alpha\}|\hat{a}_{\xi_0}\hat{a}_{\xi}^\dag \ket{\{ \alpha \}}|^2.
\end{equation}
Using the techniques in Section III, it is straightforward to show that (see Appendix~\ref{appendix:fid})
\begin{equation}
F=\frac{|\sigma(\tau)|^2(1+n_\alpha)}{n_\alpha(1+|\sigma(\tau)|^2)},
\label{eqn:fidelity}
\end{equation}
For perfect overlap of the single photon, $|\sigma(\tau)|^2=n_\alpha$, which gives $F=1$. As the overlap decreases, $|\sigma(\tau)|^2 \to 0$, and we have that $F \to 0$. In general, for a given photon number of the coherent state, $n_\alpha$, the fidelity decays with a $|\sigma(\tau)|^2/(1+|\sigma(\tau)|^2)$ dependence. The corresponding expression for the fidelity with loss present is given in Appendix~\ref{appendix:fid}. In an experiment, the fidelity could be obtained by carrying out time-dependent state tomography, such as in a reconstruction of the Wigner function for the state~\cite{Lvovsky09,MacRae12,Morin13,Ogawa16}.

As an example from Section III, for $n_\alpha=3$ and no loss, an imperfect PAC state with $\tau=5/\Omega$ and $\Omega_1=5 \Omega$ displays sub-Poissionian behaviour with $(\Delta n)^2/\langle n \rangle \simeq 0.75$ and has a second-order correlation $g^{(2)}[0]\simeq 0.94$, and is therefore nonclassical. However, $|\sigma(\tau)|^2\sim0$ and thus the state's fidelity with the ideal PAC state is effectively zero. Choosing $\tau=1/\Omega$ and $\Omega_1=\Omega$, the sub-Poissonian behaviour and second-order correlation do not change appreciably, while $F\simeq 0.70$. 
The fidelity appears to have little correlation with these two nonclassical quantities, however, they are correlated in that as $F$ increases, the sub-Poissonian behavior and second-order correlation improve slightly, reaching their minimum ideal values, as seen in Figs.~\ref{fig3} and \ref{fig5}.

For the quadrature variance we observe a larger variation. As seen in Fig.~\ref{fig10}~(b), we have a minimum of $0.15$ below the coherent state value at $\tau=0$ and $\Omega_1=\Omega$ (where $F=1$) demonstrating squeezing, which increases to the coherent state value at $\tau=1/\Omega$ and $\Omega_1=\Omega$ (where $F=0.70$) corresponding to no squeezing. Thus, squeezing and fidelity appear to be more correlated with each other, which can be seen more clearly by comparing the functional forms of Eqs. (\ref{eqn:Xphiminvar_pacs}) and (\ref{eqn:fidelity}). Further work on connecting continuous-mode fidelity to quadrature squeezing and nonclassicality in general is an interesting direction for future studies. 

\section{Conclusion} 
In this work we used a continuous-mode formalism for PAC states to describe how various properties are affected by timing and bandwidth imperfections. We also included loss during propagation. The properties studied were the photon-number distribution, the second-order correlations, quadrature squeezing and fidelity. For the photon-number distribution we calculated its mean and variance, and used these to quantify the degree of nonclassicality in terms of how much the variance to mean ratio went below unity, where the state becomes sub-Poissonian. We found the ratio to be robust to temporal and spectral mismatch, and that for increasing loss the ratio had a linear dependence. For the second-order correlations, we found that they are also robust to temporal and spectral mismatch, with values consistently in the quantum (nonclassical) regime and unaffected by loss. For the quadratures, we found that the variance is again robust to temporal and spectral mismatch, although not as much as the photon-number distribution and correlations. Squeezing was found for a range of temporal and spectral mismatch, further highlighting the nonclassical nature of the continuous-mode PAC state. For the fidelity, we derived the functional dependence describing the closeness of a PAC state to the ideal case and found a small correlation between the fidelity and the sub-Poissonian behavior, as well as the second-order correlation. On the other hand, the fidelity and quadrature variance had an improved correlation.

The combined results of this study may aid the further development of robust schemes for PAC state generation and their use in quantum information applications. Future work could study multi-photon added coherent states and quantify the performance of continuous-mode PAC states in quantum sensing.

{\it Acknowledgements.---} This research was supported by the South African National Research Foundation, the National Laser Centre and the South African Research Chair Initiative of the Department of Science and Innovation and National Research Foundation.

\appendix


\section{Continuous mode formalism}
Here we provide some brief details of continuous-mode pulsed number states and coherent states. These details are needed to derive the photon number distribution, second-order correlation function, and the mean and variance of the quadrature operator for each state, including the PAC state.
\label{appendix: CMF}

\subsection{Photon number states}
A number state in a single spatial mode, with $n$ photons arbitrarily distributed in time, may be expressed as \cite{Loudon00,Ou06}
\begin{equation}
	\ket{{n}_{\beta}} =\int dt_1 ...dt_n \beta(t_1,...,t_n) \ket{\{1_{t_i}\}_{i=1}^{n}},
	\label{eqn:nbeta}
\end{equation}
where $\beta(t_1,\cdots,t_n)$ is an $n$-fold joint wavefunction. Here we use a transient number state basis, $\ket{\{1_{t_i}\}_{i=1}^{n}}$, defined in Eq.~\eqref{eqn:nt},
which has the orthonormality relation
\begin{equation}
	\braket{\{1_{t'_i}\}_{i=1}^{m}|\{1_{t_i}\}_{i=1}^{n}} = \frac{1}{{\cal N}} \delta_{n,m} \sum_{\forall \rho} \prod_{i=1}^{n}\delta(t'_i - t_{\rho(i)}).
	\label{eqn:ntortho}
\end{equation}
The normalisation is given by ${\cal N}=\int dt_1\cdots dt_n \beta^*(t_1,\cdots,t_n)\sum_{\forall \rho}\beta(t_{\rho(1)},\cdots,t_{\rho (n)})$. The summation is over all permutations, $\rho$, of the sequence $\{1,\cdots,n\}$. For overlapping photons which are indistinguishable, $\beta(t_1,\cdots,t_n)$ is symmetric under pair-wise permutations of its time arguments and $\int dt_1\cdots dt_n |\beta(t_1,\cdots,t_n)|^2=1$. In this case, the normalisation is simply ${\cal N}=n!$. For an arbitrary state, some of the $n$ photons may be distinguishable. In this case $\beta(t_1,\cdots,t_n)$ is partially symmetric and may be factored into a product of fully symmetric wavefunctions~\cite{Ou06}. The transient number state normalisation, ${\cal N}$, would then change as a result. One may also fix the normalisation of the transient number state in Eq.~(\ref{eqn:nbeta}) to be $n!$ and rescale the partially symmetric wavefunction $\beta$ so that it satisfies the relation $\int dt_1\cdots dt_n \beta^*(t_1,\cdots,t_n)\sum_{\forall \rho}\beta(t_{\rho(1)},\cdots,t_{\rho (n)})=n!$. Thus, in what follows we fix the normalisation of the transient number states to that of the symmetric case, ${\cal N}=n!$.

We use the transient number state basis for an arbitrary state as it is more convenient when calculating the photon number probability density. We will also use the above relations to derive various formulas. However, for the moment we will consider the special case of number states of $n$ independent and indistinguishable photons. For such a state we may write $\beta(t_1,...,t_n) = \xi(t_1)...\xi(t_n)$, thereby reducing Eq. \eqref{eqn:nbeta} to
\begin{equation}
	\ket{n_\xi} = \frac{1}{\sqrt{n!}} (\axid)^n \ket{0}.
	\label{eqn:nxi}
\end{equation}

The photon wavepacket creation operator $\axid$, given by Eq.~\eqref{eqn:axid}, creates a photon with a temporal wavepacket amplitude $\xi(t)$ that is normalised;
\begin{equation}
	\int dt |\xi(t)|^2 = 1.
	\label{eqn:modsqdxi}
\end{equation}
The Hermitian conjugate, $\axi$, absorbs a single-photon wavepacket, and with $\axid$ obeys the commutation relation
\begin{equation}
	[\axi,\axid] = 1.
	\label{eqn:axicom} 
\end{equation}

Additionally, we have $[\at,\axid] = \xi(t)$, from which one may derive the more general relations
\begin{equation}
	\Bigg[\prod_{i=1}^{n}\at[i],\axid\Bigg] = \sum_{k=1}^{n}\Bigg(\xi(t_k)\prod_{\substack{i=1\\i\neq k}}^{n}\hat{a}(t_i)\Bigg)
	\label{eqn:atnaxi} 
\end{equation}
and
\begin{equation}
	\Big[\at,\big(\axid\big)^n\Big] =n \xi(t)\big(\hat{a}_\xi^\dagger\big)^{n-1}.
	\label{eqn:ataxin} 
\end{equation}
From the commutation relation in Eq.~\eqref{eqn:ataxin} and the number state definition in Eq.~\eqref{eqn:nxi}, it follows that
\begin{equation}
	\hat{a}(t)\ket{n_\xi} = \sqrt{n}\xi(t)\ket{(n-1)_\xi}.
	\label{eqn:at_ns}
\end{equation}

\subsection{Coherent state}
A continuous-mode coherent state in a single spatial mode is defined as $\coh=  \exp(\aalphad - \aalpha)\ket{0}$, where the creation operator, $\aalphad$, is defined similarly to $\axid$ in Eq.~\eqref{eqn:axid}, but with $\xi(t) \rightarrow \alpha(t)$~\cite{Loudon00}. Unlike the function $\xi(t)$, $\alpha(t)$ is not normalised but is square-integrable;
\begin{equation}
	\int dt |\alpha(t)|^2 = n_\alpha.
	\label{eqn:modsqdalpha}
\end{equation}
The definition of the continuous-mode coherent state is not explicitly used, however, we make use of the eigen-relation,
\begin{equation}
	\at \ket{\{\alpha\}} = \alpha(t)\ket{\{\alpha\}},
	\label{eqn:eigen}
\end{equation}
when calculating expectation values.


\section{Photon statistics}
\label{appendix:P}

\subsection{Photon number distribution}

\subsubsection{Coherent state}
\label{appendix:pCS}
The photon number distribution of a coherent state is derived in the same manner as that of the PAC state. The transient number state is expanded in operator form by using the Hermitian conjugate of Eq.~\eqref{eqn:nt} in Eq.~\eqref{eqn:Pdensity}. The eigen-relation in Eq.~\eqref{eqn:eigen} is then used to carry out the operation of each $\hat{a}(t_i)$ on $\ket{\psi}=\ket{\{\alpha\}}$. The resulting probability density is
\begin{equation}
\begin{split}
	P_n\big(\{t_i\}_{i=1}^{n}\big) = & ~|\langle\{1_{t_i}\}_{i=1}^{n}\coh|^2\\
	= & ~\frac{1 }{n!}~\mathrm{e}^{-n_\alpha}~\prod_{i=1}^{n}|\alpha(t_i)|^2.						
\end{split}
\label{eqn:Pdensity_cs}	
\end{equation}
Integrating Eq.~\eqref{eqn:Pdensity_cs}, gives the lossless photon distribution
\begin{equation}
\begin{split}
	P_n = & ~\frac{\mathrm{e}^{-n_\alpha}}{n!}\prod_{i=1}^{n}\int dt_i~|\alpha(t_i)|^2	\\ = & ~\frac{{n_\alpha}^n}{n!}~\mathrm{e}^{-n_\alpha},
\end{split}	
\label{eqn:Pn_cs}
\end{equation}
which is equal to the usual single-mode result~\cite{Loudon00}, and is valid for $n = 0, 1, 2, \cdots$. Substituting this result into Eq.~\eqref{eqn:PnL} introduces the effects of loss, giving
\begin{equation}
	P_n(L) = {|\eta(L)|}^n~\frac{n_\alpha^n}{n!}\mathrm{e}^{-|\eta(L)|n_\alpha}.
	\label{eqn:PnL_cs}
\end{equation}
This distribution has a mean of $\langle n \rangle = |\eta(L)| n_\alpha$ and a variance of
${(\Delta n)}^2 = |\eta(L)| n_\alpha$, confirming that the wavepacket state continues to have Poissonian statistics as it propagates.

\subsubsection{Number state}
\label{appendix:pNS}
The photon number probability density for a number state $\ket{m_\beta}$, is obtained by using the number state definition in Eq.~\eqref{eqn:nbeta} and substituting it for $\ket{\psi}$ in Eq.~\eqref{eqn:Pdensity}. 
\begin{equation}
	\begin{split}
		P_n\big(\{t_i\}_{i=1}^{n}\big)  = & ~|\braket{\{1_{t_i}\}_{i=1}^{n}|m_\beta}|^2\\
		= & ~\Bigg|\int dt'_1\cdots dt'_m ~\beta(t'_1,\cdots,t'_m)~\braket{\{1_{t_i}\}_{i=1}^{n}|\{1_{t'_j}\}_{j=1}^{m}}\Bigg|^2.\\
	\end{split}
\end{equation}
Using the orthonormality relation in Eq.~\eqref{eqn:ntortho} leads to
\begin{equation}
\begin{split}
	P_n\big(\{t_i\}_{i=1}^{n}\big)  = & ~\Bigg|\int dt'_1\cdots dt'_m ~\beta(t'_1,\cdots,t'_m)\\
	&\times~\sum_{\forall \rho}\prod_{i=1}^{n}\delta(t'_i - t_{\rho(i)})\delta_{n,m}/n!\Bigg|^2\\
	= & ~\Big|\sum_{\forall \rho}\beta(t_{\rho(1)},\cdots,t_{\rho(m)})\delta_{n,m}/n!\Big|^2.
\end{split}
\end{equation}
Setting $m=n$ due to $\delta_{n,m}$, we have that for $n$ indistinguishable photons, $\beta(t_{\rho(1)},\cdots,t_{\rho(n)})$ is symmetric with respect to the interchange of any pair of times and so we may choose any permutation, {\it e.g.} $\rho(i)=i$. We can then write $\sum_{\forall \rho}\beta(t_{\rho(1)},\cdots,t_{\rho(n)}) = n!\beta(t_1,\cdots,t_n)$, where $n!$ is the total number of permutations. The probability density is then
\begin{equation}
\begin{split}
P_n\big(\{t_i\}_{i=1}^{n}\big)  = & |\beta(t_1,\cdots,t_n)|^2\delta_{n,m}.
\end{split}
\end{equation}
Integrating with respect to all times, we obtain the distribution
\begin{equation}
	\begin{split}
		P_n  = & \int dt_1\cdots dt_n |\beta(t_1,\cdots,t_n)|^{2}~\delta_{n,m}=\delta_{n,m},
	\end{split}
	\label{eqn:Pntnumber}
\end{equation}
which is the expected result as there are exactly $m$ photons in the state. Using Eq.~\eqref{eqn:P0}, the vacuum probability is found to be  $P_0 = \delta_{0,m}$. Thus Eq.~\eqref{eqn:Pntnumber} is valid for $n=0,1,2,\cdots$. With the lossless photon population known, the case with loss is found by substituting Eq.~\eqref{eqn:Pntnumber} into Eq.~\eqref{eqn:PnL}, giving
\begin{equation}
	P_n(L) = {|\eta(L)|}^n~{|1 - \eta(L)|}^{m-n}~\binom{m}{n},
	\label{eqn:PnLnumber}
\end{equation}
which coincides with the single-mode result. 

The mean of this distribution is given by $\langle n \rangle = m|\eta(L)|$ and the variance by ${(\Delta n)}^2 =  m~|\eta(L)|~\Big(1-|\eta(L)|\Big)$. Single-mode number states are known to have sub-Poissonian statistics~\cite{Loudon00}. Taking the ratio of the mean and variance, we obtain $(\Delta n)^2/{\langle n\rangle} = 1 -  |\eta(L)| \leq 1$, thus confirming the sub-Poissonian nature of number state wavepackets.

\subsection{Second-order correlation function}

\subsubsection{Coherent state}
As was done for the PAC state, we use the eigen-relation in Eq.~\eqref{eqn:eigen} to evaluate the averages. Unlike the PAC state normal-ordering is not necessary. The photon flux is
\begin{equation}
\begin{split}
f_1(t) =&~ \braket{\{\alpha\}|\hat{a}^\dagger(t)\hat{a}(t)|\{\alpha\}}=|\alpha(t)|^2,
\end{split}
\label{eqn:f1_cs}
\end{equation}
which yields a total photon number equal to the mean, $n_\alpha$, upon integration. We similarly find the coincidence rate
\begin{equation}
\begin{split}
f_2(t_1, t_2) =&~ \braket{\{\alpha\}|\hat{a}^\dagger(t_1)\hat{a}^\dagger(t_2)\hat{a}(t_2)\hat{a}(t_1)|\{\alpha\}}\\
=&~~|\alpha(t_1)|^2|\alpha(t_2)|^2.
\end{split}
\label{f2_cs}
\end{equation}
Using these expressions in Eq.~\eqref{eqn:g2t}, one obtains the second-order correlation function~\cite{Loudon00},
\begin{equation}
\begin{split}
g^{(2)}(t_1,t_2) =&~ \frac{|\alpha(t_1)|^2|\alpha(t_2)|^2}{|\alpha(t_1)|^2|\alpha(t_2)|^2}=1.
\end{split}
\label{eqn:g2_cs}
\end{equation}
Using the expressions for $f_1$ and $f_2$ it is straightforward to show that $g^{(2)}[0]=1$.
\subsubsection{Number state}
The photon flux for the number state is calculated by applying $\hat{a}(t)$ to $\ket{n_\xi}$ using Eq.~\eqref{eqn:at_ns}, and multiplying this by its Hermitian conjugate, which gives
\begin{equation}
\begin{split}
f_1(t) =&~ \braket{n_\xi|\hat{a}^\dagger(t)\hat{a}(t)|n_\xi}=n|\xi(t)|^2.
\end{split}
\label{eqn:f1_ns}
\end{equation}
Similarly we obtain the coincidence flux,
\begin{equation}
\begin{split}
f_2(t_1, t_2) =&~ \braket{n_\xi|\hat{a}^\dagger(t_1)\hat{a}^\dagger(t_2)\hat{a}(t_2)\hat{a}(t_1)|n_\xi}\\
=&~~n(n-1)|\xi(t_1)|^2|\xi(t_2)|^2.
\end{split}
\label{eqn:f2_ns}
\end{equation}
Using Eq.~\eqref{eqn:f1_ns} and \eqref{eqn:f2_ns} in Eq.~\eqref{eqn:g2t} yields a time-independent second-order correlation function,
\begin{equation}
\begin{split}
g^{(2)}(t_1,t_2) =&~ \frac{f_2(t_1,t_2)}{f_1(t_1)f_1(t_2)}\\
=&~ \frac{n(n-1)|\xi(t_1)|^2|\xi(t_2)|^2}{n^2|\xi(t_1)|^2|\xi(t_2)|^2}\\
=&~ 1 - \frac{1}{n}\\
=&~ g^{(2)}[0].
\end{split}
\label{eqn:g2_ns}
\end{equation}
These results are equal to the single-mode case~\cite{Loudon00}.

\section{Quadratures}
\label{appendix:Q}

\subsection{Coherent state}
The coherent state is well known as a minimum-uncertainty state; the uncertainty product, $\Delta X_\varphi(t, T)\Delta X_{\varphi + \pi/2}(t, T)$, assumes the minimum permissible value. As shown below, a coherent state wavepacket has a constant quadrature variance over all $\varphi$, thus making it non-squeezed. Any state with one quadrature variance less than that of the coherent state value is considered squeezed~\cite{Loudon00}.

We begin by evaluating Eq.~\eqref{eqn:Xphi_mean} for $\ket{\{\alpha\}}$, with the help of the eigen-relation in Eq.~\eqref{eqn:eigen}, which gives the mean
	\begin{equation}
		\braket{\hat{X}_\varphi(t, T)} = \int_{t}^{t + T} dt'~\mathrm{Re}\Big\{\alpha(t') \mathrm{e}^{-\im \varphi(t')}\Big\}.
		\label{eqn:Xphimeancoh}
	\end{equation}
It can similarly be shown that
	\begin{equation}
	\braket{:\hat{X}_\varphi^2(t, T):} = \braket{\hat{X}_\varphi(t, T)}^2,
	\end{equation}
which cancels the square of Eq.~\eqref{eqn:Xphimeancoh} when both are substituted into Eq.~\eqref{eqn:Xphi_var} to calculate the variance. This removes the dependence on $\varphi$ and the wavepacket profile, giving
	\begin{equation}
	\Big(\Delta X_{\varphi}(t, T)\Big)^2 = \frac{T}{4}.
	\end{equation}
This result differs from the single-mode value of $1/4$ only by the factor $T$, which occurs due to the integration. The vacuum state shares the same quadrature variance.
\newline

\label{appendix:qCS}

\subsection{Number state}
Due to the orthogonality of number states, the quadrature mean in Eq.~\eqref{eqn:XphiL_mean} evaluates to
	\begin{equation}
	\braket{\hat{X}_\varphi(t, T)} = 0
	\label{eqn:Xmean_ns}
	\end{equation}
when taking the  expectation value with respect to $\ket{n_\xi}$. Similarly, only the second term of Eq.~\eqref{eqn:Xphi2moment} with $\hat{a}^{\dagger}(t'')\hat{a}(t')$ remains, which simplifies to
	\begin{equation}
	\begin{split}
	\braket{:{\hat{X}_\varphi}^2(t, T):} = & \frac{n}{2}\Bigg[\int_{t}^{t + T}dt'~|\xi(t')|\cos\Big(\theta_1(t') - \varphi(t')\Big)\Bigg]^2\\
	& +\frac{n}{2}\Bigg[\int_{t}^{t + T}dt'~|\xi(t')|\sin\Big(\theta_1(t') - \varphi(t')\Big)\Bigg]^2.
	\end{split}
	\label{eqn:X2mean_ns}
	\end{equation}
When $\varphi(t') \neq \theta_1(t')$, the integrands oscillate about zero with an envelope of $|\xi(t')|$. For $T \gg 1/\Omega_1$, the integral evaluates to roughly zero. Setting $\varphi(t') = \theta_1(t)$ removes the oscillations and the integrals evaluate to a positive value. Thus, only the quadrature in phase with the field produces a non-trivial result,
	\begin{equation}
	\begin{split}
	\braket{:{\hat{X}_{\theta_1}}^2(t, T):} = &\frac{n}{2}\Bigg[\int_{t}^{t + T} dt'~|\xi(t')|\Bigg]^2\\
	\leq&\frac{nT}{2}\int_{t}^{t + T} dt'~{|\xi(t')|}^2.
	\end{split}
	\end{equation}
The upper bound is obtained by applying the Cauchy-Schwartz inequality to the squared integral. For $T\gg1/\Omega_1$, the upper bound is approximately $nT/2$ which resembles the single-mode result. Equality is achieved when $|\xi(t')|$ is constant in time, in which case $\int_{-\infty}^{+\infty}dt|\xi(t)|^2$ is no longer finite, making a continuous-mode treatment unsuitable. For this reason, we leave the upper bound of the variance open. Using Eqs.~\eqref{eqn:Xmean_ns} and \eqref{eqn:X2mean_ns} in Eq.~\eqref{eqn:XphiL_var} gives the quadrature variance,
	\begin{equation}
	\Big(\Delta X_{\theta_1}(t, T)\Big)^2 = \frac{T}{4} + \frac{n}{2}\Bigg[\int_{t}^{t + T} dt'~|\xi(t')|\Bigg]^2,
	\end{equation}
which lies between $T/4$ and $(2n+1)T/4$.

\label{appendix:qNS}

\section{Fidelity}
In the main text the expression for the fidelity is given by 
\begin{equation}
F=|\langle \psi | \phi \rangle|^2=\bra{\psi} \hat{F}_\phi \ket{\psi},
\end{equation}
\newline
where $\ket{\psi}=\ket{\{ \alpha \},1_{\xi_0}}_g\ket{0}_e=\frac{1}{(1+n_\alpha)^{1/2}}\hat{a}_{\xi_0}\ket{\{ \alpha \}}_g \ket{0}_e$ is the ideal state and $\hat{F}_\phi=\ket{\phi}\bra{\phi}$ is the fidelity operator for the imperfect state. We have included the environment mode ($e$) in addition to the guided mode ($g$) to account for loss during propagation. In the Heisenberg picture, the ideal state remains fixed while the fidelity operator (the observable) is modified. Under the transformation in Eq.~(\ref{eqn:aL}) we have $\ket{\phi}\to \ket{\phi'}$ with
\begin{eqnarray}
\ket{\phi'} &=& |N(\tau)|^{1/2} \int dt' \xi(t'+\tau)[\eta^{1/2} \hat{a}^\dag (t') - i (1-\eta)^{1/2}\hat{v}^\dag(t') ] \nonumber \\ 
&& \qquad \times \ket{\{ \eta^{1/2} \alpha \}}_g\ket{\{ -i(1-\eta)^{1/2} \alpha \} }_e.
\end{eqnarray}
Here, the $L$ dependence of $\eta$ has been suppressed for convenience and the retarded time is set to the non-retarded time (equivalent to the ideal state shifted in time to account for propagation without loss). We then have
\begin{eqnarray}
\langle \psi| \phi' \rangle &=& \eta^{1/2} \frac{|N(\tau)|^{1/2}}{(1+n_\alpha)^{1/2}}  \langle 0| \{ -i(1-\eta)^{1/2} \alpha \} \rangle_e \langle \{ \alpha \}| \{ \eta^{1/2} \alpha \} \rangle_g  \nonumber \\ 
&& \qquad \times [\sigma^*(\tau)/\sqrt{n_\alpha}+\eta^{1/2} \sqrt{n_\alpha}\sigma^*(\tau)].
\end{eqnarray}
One can then use $\langle \{ \beta\} |\{\alpha \}\rangle=e^{-n_\beta/2} \bra{0} \sum_{n=0}^{\infty}\frac{(\hat{a}_\beta)^n}{n!}\ket{\{ \alpha\}}=e^{-n_\beta/2}e^{-n_\alpha/2}e^{\mu}$, where $\mu=\int dt \beta^*(t)\alpha(t)$. This leads to 
\begin{equation}
F'=|\langle \psi | \phi' \rangle|^2=\eta e^{-2n_\alpha(1-\eta^{1/2})} \frac{|N(\tau)|}{(1+n_\alpha)}\frac{|\sigma(\tau)|^2}{n_\alpha}(1+\eta^{1/2}n_\alpha)^2,
\end{equation}
with $\mu=\eta^{1/2} n_\alpha$. For $\eta=1$ we recover the result in the main text for the case of no loss.

\label{appendix:fid}



\begin{thebibliography}{99}

\bibitem{Dodonov02} V. V. Dodonov, J. Opt. B: Quant. Semiclass. Opt. {\bf 4}, R1 (2002).

\bibitem{Braunstein05} S. L. Braunstein and P. van Loock, Rev. Mod. Phys. {\bf 77}, 513 (2005).

\bibitem{Waks06} E. Waks, E. Diamanti and Y. Yamamoto, New J. Phys. {\bf 8}, 4 (2006).

\bibitem{Satya85} M. V. Satyanarayana, Phys. Rev. D {\bf 32}, 400 (1985).

\bibitem{Ziesel13} F. Ziesel, T. Ruster, A. Walther, H. Kaufmann, S. Dawkins, K. Singer, F. Schmidt-Kaler and U. Poschinger, J. Phys. B: Atom. Mol. Opt. Phys. {\bf 46}, 104008 (2013).

\bibitem{Lvovsky15} A. I. Lvovsky, {\sl Squeezed Light}, Photonics: Scientific Foundations, Technology and Applications, Volume 1, pp. 121-163 (Wiley Online, 2015).

\bibitem{Agarwal91} G. S. Agarwal and K. Tara, Phys. Rev. A {\bf 43}, 492 (1991).

\bibitem{Agarwal92} G. S. Agarwal and K. Tara, Phys. Rev. A {\bf 46}, 485 (1992).

\bibitem{Sivakumar99} S. Sivakumar, J. Phys. A: Math. Gen. {\bf 32}, 3441 (1999).

\bibitem{Sivakumar00} S. Sivakumar, J. Opt. B: Quant. Semiclass. Opt. {\bf 2}, R61 (2000).

\bibitem{Gard16} B. T. Gard {\it et al.}, arXiv:1606.09598v2 (2016).

\bibitem{Braun14} D. Braun, P. Jian, O. Pinel and N. Treps, Phys. Rev. A {\bf 90}, 013821 (2014).

\bibitem{Schnabel17} R. Schnabel, Phys. Rep. {\bf 684}, 1-51 (2017).

\bibitem{Loepp06} S. Loepp and W. K. Wootters, Protecting Information (Cambridge University, Cambridge, England, 2006).

\bibitem{Assche06} G. Van Assche, Quantum Cryptography and Secret-Key Distillation (Cambridge University, Cambridge, England, 2006).

\bibitem{Barnett06} S. M. Barnett, Quantum Information (Oxford University, Oxford, 2009).

\bibitem{Barnett18} S. M. Barnett, G. Ferenczi, C. R. Gilson and F. C. Speirits, Phys. Rev. A {\bf 98}, 013809 (2018).

\bibitem{Kim08} M. S. Kim, J. Phys. B: At. Mol. Opt. Phys. {\bf 41}, 13 (2008).

\bibitem{Fiurasek09} J. Fiur\'a\v{s}ek, Phys. Rev A {\bf 80}, 053822 (2009).

\bibitem{Parigi07} V. Parigi, A. Zavatta, M. S. Kim and M. Bellini, Science {\bf 317}, 1890 (2007).

\bibitem{Vidrighin16} M. D. Vidrighin, O. Dahlsten, M. Barbieri, M. S. Kim, V. Vedral and I. A. Walmsley, Phys. Rev. Lett. {\bf 116}, 050401 (2016).

\bibitem{RamosPrieto14} I. Ramos-Prieto, B. M. Rodriguez-Lara and H. M. Moya-Cessa, Int. J. Quantum Inf. {\bf 12}, 1560005 (2014).

\bibitem{Mojaveri14} B. Mojaveri and A. Dehghani, Eur. Phys. J. D {\bf 68}, 315 (2014).

\bibitem{Shringapure19} S. U. Shringarpure and J. D. Franson, Phys. Rev. A {\bf 100}, 043802 (2019).

\bibitem{Kalamidas08} D. Kalamidas, C. C. Gerry and A. Benmoussa, Phys.
Lett. A {\bf 372}, 1937 (2008).

\bibitem{Li07} Y. Li, H. Jing and M. S. Zhan, Chin. Phys. {\bf 16}, 1883 (2007).

\bibitem{Ren19} G. Ren and W.-H. Zhang, J. Mod. Opt. {\bf 66}, 1408 (2019).

\bibitem{Dominguez16} F. A. Dom\'{\i}nguez-Serna, F. J. Mendieta-Jimenez and F. Rojas, Quant. Inf. Proc. {\bf 15}, 3121 (2016).

\bibitem{Hu09} L.-Y. Hu and H.-Y. Fan, Phys. Scr. {\bf 79} 035004 (2009).

\bibitem{Filippov13} S. N. Filippov, V. I. Manko, A. S. Coelho, A. Zavatta and M. Bellini, Phys. Scr. {\bf T153}, 014025 (2013).

\bibitem{HongChun10} Y. Hong-Chun, X. Xue-Xiang and F. Hong-Yi, Chin. Phys. B {\bf 19}, 104205 (2010).

\bibitem{Sivakumar13} S. Sivakumar, Int. J. Theor. Phys. {\bf 53}, 1697 (2014).

\bibitem{Zavatta04} A. Zavatta, S. Viciani and M. Bellini, Science {\bf 306}, 660 (2004).

\bibitem{Zavatta05} A. Zavatta, S. Viciani and M. Bellini, Phys. Rev. A {\bf 72}, 023820 (2005).

\bibitem{Barbieri10} M. Barbieri, N. Spagnolo, M. G. Genoni, F. Ferreyrol, R. Blandino, M. G. A. Paris, P. Grangier and R. Tualle-Brouri, Phys. Rev. A {\bf 82}, 063833 (2010).

\bibitem{Blow90} K. J. Blow, R. Loudon, S. J. D. Phoenix and T. J. Shepherd, Phys. Rev. A {\bf 42}, 4102 (1990).

\bibitem{Loudon00} R. Loudon, {\sl The Quantum Theory of Light}, 3$^{\rm rd}$ Ed. (Oxford University Press, Oxford, 2000). 

\bibitem{Vahala93} K. J. Vahala, Pure Appl. Opt. {\bf 2}, 549-60 (1993).

\bibitem{Fischer18} K. A. Fischer, Continuum {\bf 1}, 772 (2018).

\bibitem{Resch02} K. J. Resch, J. S. Lundeen and A. M. Steinberg, Phys. Rev. Lett. {\bf 88},113601 (2002).

\bibitem{Rarity05} J. G. Rarity, P. R. Tapster and R. Loudon, J. Opt. B: Quant. Semiclass. Opt. {\bf 7}, S171 (2005).

\bibitem{Akimov07} A. V. Akimov, A. Mukherjee, C. L. Yu, D. E. Chang, A. S. Zibrov, P. R. Hemmer, H. Park and M. D. Lukin, Nature {\bf 450}, 402-406 (2007).

\bibitem{DiMartino12} G. Di Martino, Y. Sonnefraud, S. K\'{e}na-Cohen, M. S. Tame, \c{S}. K. \"{O}zdemir, M. S. Kim and S. A. Maier, Nano Lett. {\bf 12}, 2504-2508 (2012).

\bibitem{OBrien05} J. O'Brien, A. Furusawa and J. Vu\v{c}kovi\'c, Nat. Photon. {\bf 3}, 687-695 (2005).

\bibitem{Tame08} M. S. Tame, C. Lee, J. Lee, D. Ballester, M. Paternostro, A. V. Zayats and M. S. Kim, Phys. Rev. Lett. {\bf 101}, 190504 (2008).

\bibitem{Ou06} Z. Y. Ou, Phys. Rev. A {\bf 74}, 063808 (2006).

\bibitem{Jeffers93} J. R. Jeffers, N. Imoto and R. Loudon, Phys. Rev. A {\bf 47}, 3346 (1993).

\bibitem{Gardiner00} C. W. Gardiner and P. Zoller, {\sl Quantum noise}, 2$^{\rm nd}$ Ed. (Springer-Verlag, Berlin, 2000). 

\bibitem{Fischer16} K. A. Fischer, K. M\"uller, K. G. Lagoudakis and J. Vu\v{c}kovi\'c, New J. Phys. {\bf 18}, 113053 (2016).

\bibitem{Tan19} K. C. Tan and H. Jeong, Quantum Rep. {\bf 1}, 151 (2019).

\bibitem{Knott16} P. A. Knott, T. J. Proctor, A. J. Hayes, J. P. Cooling and J. A. Dunningham, Phys. Rev. A {\bf 93}, 033859 (2016).

\bibitem{Lvovsky09} A. I. Lvovsky and M. G. Raymer, Rev. Mod. Phys. {\bf 81}, 299 (2009).

\bibitem{Ogawa16} H. Ogawa, H. Ohdan, K. Miyata, M. Taguchi, K. Makino, H. Yonezawa, J. I. Yoshikawa and A. Furusawa, Phys. Rev. Lett. {\bf 116}, 233602 (2016).

\bibitem{Nielsen10} M. A. Nielsen and I. L. Chuang, Quantum Computation and Quantum Information: 10th Anniversary Edition (Cambridge University Press, Cambridge, 2010). 

\bibitem{MacRae12} A. MacRae, T. Brannan, R. Achal and A. I. Lvovsky, Phys. Rev. Lett. {\bf 109}, 033601 (2012).

\bibitem{Morin13} O. Morin, C. Fabre and J. Laurat, Phys. Rev. Lett. {\bf 111}, 213602 (2013).

\bibitem{Ogawa16} H. Ogawa, H. Ohdan, K. Miyata, M. Taguchi, K. Makino, H. Yonezawa, J. Yoshikawa and A. Furusawa, Phys. Rev. Lett. {\bf 116}, 233602 (2016).

\end{thebibliography}
\end{document}